%% file: bound_avg.tex
\documentclass[aps,prl,twocolumn,groupedaddress,showpacs,superscriptaddress,amssymb,amsmath]{revtex4-2}
\usepackage{graphicx}
\usepackage{amsmath}
\usepackage{amssymb}
\usepackage{amsfonts}
\usepackage{color}
\usepackage{bm}        % for math
\usepackage{comment}
\usepackage{placeins}
\usepackage[mathscr]{eucal}
\usepackage[colorlinks, linkcolor=red,citecolor=blue,urlcolor=blue]{hyperref}
\usepackage[normalem]{ulem}
\usepackage[all]{hypcap}
\usepackage{multirow}
\usepackage[dvipsnames,usenames,table]{xcolor}
\usepackage{dcolumn}
\usepackage{hyperref}
\usepackage{bm}
\usepackage{epsf}
\usepackage{subfigure}
\usepackage{amsfonts}
\usepackage{epstopdf}%
\newcommand{\be}{\begin{equation}}
\newcommand{\ee}{\end{equation}}
\newcommand{\bea}{\begin{eqnarray}}
\newcommand{\eea}{\end{eqnarray}}
\begin{document}
\title{Universal Bounds on Fluctuations in Continuous Thermal Machines} %Heat Engines}
%\title{Universal Lower Bounds for Heat Engine Fluctuations}
\author{Sushant Saryal}
\affiliation{Department of Physics,
		Indian Institute of Science Education and Research, Pune 411008, India}
\author{Matthew Gerry}
\affiliation{Department of Physics, University of Toronto, Toronto, Ontario, Canada M5S 1A7}	
	\author{Ilia Khait}
\affiliation{Department of Physics, University of Toronto, Toronto, Ontario, Canada M5S 1A7}		
	\author{Dvira Segal}
\affiliation{Department of Chemistry and Centre for Quantum Information and Quantum Control,
University of Toronto, 80 Saint George St., Toronto, Ontario, M5S 3H6, Canada}
\affiliation{Department of Physics, University of Toronto, Toronto, Ontario, Canada M5S 1A7}
		\email{dvira.segal@utoronto.ca}
\author {Bijay Kumar Agarwalla}
\affiliation{Department of Physics,
		Indian Institute of Science Education and Research, Pune 411008, India}
		\email{bijay@iiserpune.ac.in}
		
\date{\today}

\begin{abstract}
We study bounds on ratios of fluctuations in steady-state time-reversal heat engines
%driven  DDD: driven sounds to me like with time dependence
controlled by multi affinities. In the linear response regime, we prove that the relative fluctuations
(precision) of the output current (power) is always lower-bounded by the relative fluctuations of the input current (heat current absorbed from the hot bath).
As a consequence, the ratio between the fluctuations of the output and input currents
are bounded both from above and below, where the lower (upper) bound is determined
by the square of the averaged efficiency (square of the Carnot efficiency) of the engine.
The saturation of the lower bound is achieved in the tight-coupling limit
when the determinant of the Onsager response matrix vanishes.
Our analysis can be applied to different operational regimes, including engines,
refrigerators, and heat pumps.
We illustrate our findings in two types of continuous engines:
two-terminal coherent thermoelectric junctions and three-terminal quantum absorption refrigerators.
Numerical simulations in the far-from-equilibrium regime suggest that these bounds apply more broadly, beyond linear response.
%For  steady-state  time-reversal  engine,  driven  by  multi  affinities,  we
%prove that in the close to equilibrium regime, the relative fluctuation (precision) of output current (power) is always lower bounded by the corresponding relative fluctuation for the input current (heat current absorbed from the hot bath).  
%As a consequence, the ratio for the output current fluctuation over the input current fluctuation receives both lower and upper bounds where the lower (upper) bound gets determined by the square of the average efficiency (square of the Carnot efficiency) of the engine. The saturation of the bound is achieved in the tight-coupling limit leading to vanishing determinant of the Onsager response matrix. Our analysis is valid for other operational regimes such as refrigerator and heat pump. We rationalise our findings for two different setups: (i) a two-terminal coherent thermo-electric setup and (ii) a three-terminal quantum absorption refrigerator. We also analyze the validity of these bounds in the far-from-equilibrium regime.
\end{abstract}

\maketitle 

{\it Introduction.--} 
%=========================================================
% 
% second law -->fluctuations
The second law of thermodynamics plays a foundational role in many branches of science
\cite{Carnot-1,Carnot-2}. It provides an upper bound on the performance of engines and prohibits the possibility of designing an ideal engine with 100\% conversion efficiency, from wasted heat to useful work \cite{Carnot-1}.
Significant interest is currently devoted to realizing efficient, small-scale engines with
classical or quantum working fluids, made from e.g. colloidal particles, few level discrete-energy systems realized in collections of spins or quantum dots, atoms or ions
\cite{eff-1,Uzdin-theory, engine-expt1, engine-expt-spin,engine-expt-spin-osc,engine-refg,ionE}.
%following fundamental quantum mechanical principles to potentially accelerate its performance than the equivalent classical heat engine \cite{eff-1,Uzdin-theory}.
%thereby surpassing the classical limits. 
%However, this has not been achieved yet. 
% Fluctuations 
Notably, small-scale engines may suffer from significant fluctuations
(thermal and possibly quantum), which cannot be ignored when analyzing their performance. 
Unlike equilibrium conditions, fluctuations in nonequilibrium systems are more difficult to quantify %analyze 
due to the failure of the fluctuation-dissipation relation \cite{kubo,fluc-diss,Maes}.
The rapidly-growing fields of stochastic and quantum thermodynamics
\cite{st-thermo1, st-thermo2, st-thermo-Broeck, Q-thermo1,Q-thermo2,Q-thermo3} have significantly enhanced
our fundamental understanding of nonequilibrium fluctuations \cite{eff-fluc1,eff-fluc2} with practical ramifications such as in the application of electronic noise to measure molecular properties  \cite{Oren, rev-noise}.
%=============================================================
% Figure 1
\begin{figure}[h]
\includegraphics[width=\columnwidth]{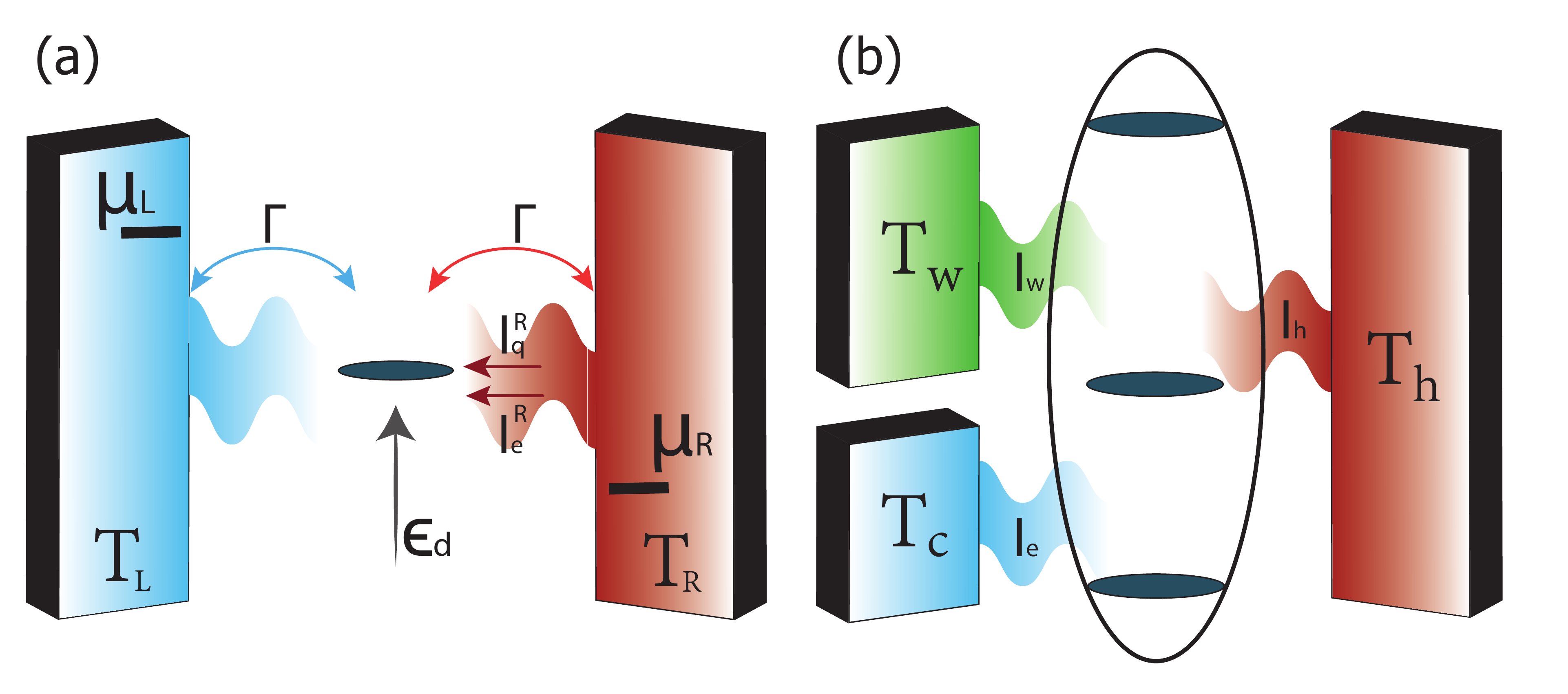}
\caption{(a) Illustration of a single-dot (energy $\epsilon_d$) thermoelectric engine with chemical potentials $\mu_{L,R}$ and temperature $T_{L,R}$. Horizontal arrows indicate the positive direction of currents in the engine, $\Gamma$ is the system-bath coupling strength. (b) Illustration of a three-level quantum absorption refrigerator; $I_{w,c,h}$ are the heat currents flowing from the baths (R,L) to the system. Other parameters are defined and visualized in Fig.~\ref{fig:QAR}. }
\label{fig:scheme}
\end{figure}
%======================================================================
% example 1: FR
The discovery of fluctuation relations \cite{st-thermo1,st-thermo2,Q-thermo2, fluc-1,fluc-2, fluc-3}
provided the underlying connection between macroscopic thermodynamics and the thermodynamics of small systems.
%pe by providing an upper bound on the mean performance in terms of Carnot efficiency. Carnot bound for heat engine efficiency, Landauer bound
%Small scale systems are often subjected to strong statistical fluctuations. For general non-equilibEquilibrium thermodynamics provides a relation between thermal fluctuation and response of a system and is known as the fluctuation-dissipation relation. However, for non-equilibrium systems, in general, such relations are not found. In the past two-decade, the discovery of fluctuation relations have provided deep understanding about statistical fluctuations in general systems.   %prQuantification of these fluctuations are of primary interest in the field of stochastic thermodynamics. 
%For example, seminal works by Seifert et al.  
%
% Example 2:TUR

Recent progress in stochastic thermodynamics concerned the derivation of bounds on relative fluctuations for observables beyond equilibrium, expressed as tradeoff relations between precision
(relative fluctuation) and cost (entropy production), now referred to as
``Thermodynamic Uncertainty relations (TUR)" \cite{Barato:2015:UncRel, trade-off-engine, Gingrich:2016:TUP,Falasco,Garrahan18,Timpanaro,Saito-TUR,Junjie-TUR, Agarwalla-TUR}.
%Polettini:2016:TUP,Pietzonka:2016:Bound,Hyeon:2017:TUR,Horowitz:2017:TUR,Proesmans:2017:TUR,Garrahan:2017:TUR,Dechant:2018:TUR,Pietzonka:2017:FiniteTUR,Falasco,SamuelssonM,Koyuk:2018:PeriodicTUR,Garrahan18,Gabri,Vu,
%Timpanaro,Hasegawa1,Hasegawa2, Saito-TUR,Junjie-TUR, Agarwalla-TUR}.
% DDD we dont need all these. Let's pick 5 most important...
The TUR further constrains the performance of a thermal engine, providing a tradeoff between
output power, power fluctuations and the engine's efficiency \cite{trade-off-engine}.
%also by the fluctuation for steady-state engines,  upper limits on . 
%has lead to the average performance of a stochastic steady-state heat engine, following Markovian dynamics, and the engines power fluctuation in terms of an upper trade-off bound.   trade-off relation that 
%an interesting trade-off relation that limits  the average performance of a steady-state heat engine following Markovian dynamics in terms of its power fluctuation. 
%for non-equilibrium currents gets lower bounded by entropy production.
%Over the past few years, various universal trade-off bounds have been discovered which are now collectively referred to as 
% Example 3: Watanabe eta-2
%The behavior of fluctuations in four-stroke heat engines were recently discussed in Ref. \cite{Watanabe}.
It was recently shown in Ref. \cite{Watanabe}
that for finite-time four-stroke heat engines, the ratio between fluctuations
of output work ($W$) and input heat ($Q$) is upper bounded, and the bound solely depends on the
temperatures of the hot $(T_h)$  and cold heat baths $(T_c)$. More precisely, the relation was given as 
$\eta^{(2)} \equiv \frac{ \langle W^2 \rangle_c }{\langle Q^2 \rangle_c} \leq \eta_C^2$,
 where $\eta_C= 1-\frac{T_c}{T_h}$ is the Carnot efficiency and $ \langle A^2 \rangle_c \!=\! \big[\langle A^2 \rangle \!-\! \langle A \rangle^2 \big]$ is the second cumulant, or the fluctuations, of an observable $A$.
 Furthermore, a tighter-than-Carnot efficiency bound was derived for classical continuous engines expressed in terms of high order fluctuations of the power \cite{Sagawa}.
 Our work exposes the relationship between these different bounds on fluctuations
 for {\it general}, quantum or classical continuous thermal machines, in linear response. % DDDD
 %
%\leq \eta_C \,\big[ \langle Q^2 \rangle \!-\! \langle Q \rangle^2 \big]^{1/2}$ 

% HERE
In this letter, we focus on generic steady-state thermal machines and prove the existence of lower and upper bounds for the ratio of output to input fluctuations, $\eta^{(2)}$.
First, we prove that in the linear response regime $\langle \eta \rangle^2 \leq \eta^{(2)} \leq \eta_C^2$,
where $\langle \eta \rangle$ is the average efficiency of the engine with $\eta_C$ as the maximal
 average efficiency.
%and $\eta^{(2)}$ is the ratio of output current  to input current fluctuations. 
These bounds follow from the Onsager reciprocity relation.
Second, we uncover the relationship between $\eta^{(2)}$ and the TUR:
For general multi-affinity systems operating as heat engines,
the relative uncertainty of the output current is always greater
than the relative uncertainty of the heat current absorbed by the engine from the hot bath.
Third, we illustrate the validity of the $\eta^{(2)}$ bounds within
the linear response regime for continuous-autonomous machines: thermoelectric junctions and quantum absorption refrigerators (QARs), depicted in Fig. \ref{fig:scheme}. Furthermore, based on simulations, we suggest the validity of our bounds in a broader regime, beyond linear response thermodynamics.
%The existence of both lower and upper bounds for fl

%. We provide a rigorous proof in the linear response regime following the Onsager's relations and 
%We furthermore provide numerical evidence to show the generality of our result.
%further making a connection to TUR. 
%This is the gap that we fill in this letter. 

%recently shown that universal upper bound for  fluctuation is put  fowa for finite-time engine, and gets bounded by the Carnot value.
%Motivated by these The motivation of our work is in fact coming from a recent study by Watanable et al  \cite{Watanabe} where the authors looked at  

%for flucuations of thermodynamic process TUR relations were shown to be valid for Markovian dynamics and it was rigorously proved for arbitrary dynamics. TUR relation states that the relative uncertainty of independent currents are bounded by the entropy production rate. 

%However, this relation doesn't give any 
%Understanding the impact of fluctuations on the mean performance of 

%{\it We need to discuss results obtained by Watanabe here.}
%we show that, in the linear response regime, following bound exists 

\noindent{\it Universal bounds in the linear response regime.--}  
 We consider a small-scale continuous engine: a system subjected to two thermodynamic forces (affinities) $A_{1}$ and $A_{2}$, which produce time-integrated stochastic currents $J_i$, $i=1,2$. % 
 The affinities  (temperature, chemical potential) are properties of the reservoirs and as thermodynamic variables they negligibly fluctuate. In contrast, the currents may suffer from significant fluctuations. 
 %, both thermal as well as quantum. What does the distinction mean here? quantum = fluctuations at T=0?
 In steady state, typically, cumulants of integrated currents scale extensively with the operation time $t$,  
 $\langle J_i^n \rangle_c = t \langle I_i^n \rangle_c$. 
 In the linear response regime, the time-intensive average currents can be expressed in terms of the Onsager response matrix \cite{review-engine,eff-stat-bijay}
%Therefore, mwe first focus on the linear response regime. In this regime, the average steady-state current can be expressed in terms of the Onsager's coefficients 
as $\langle I_{i} \rangle = \sum_{j=1,2} L_{ij} A_{j}$. The index $i=1,2$ corresponds to two different currents. $L_{ij}$ are the Onsager coefficients defined as 
%$L_{ij} = \partial_{A_{i}} J_j|_{A=0} = L_{ji}$ % DDD typos?
$L_{ij} = \partial_{A_{j}} \langle I_i\rangle|_{A=0} = L_{ji}$ 
for time-reversal systems. 
%As an example, for thermoelectric engines, $i= e, q$ corresponding to charge ($e$) and heat  ($q$) currents with the corresponding affinities $A_e=\Delta \mu/T$ and $A_q=\Delta T/T^2$  with $T=(T_h+T_c)/2$.  % DDD
As per our convention, average currents $\langle I_i \rangle$ are positive when flowing towards the system. 
% DDD Not left to right? Check that we are consistent. 
Considering now fluctuations of currents around their mean values, 
%the dispersion $D_i = \lim_{t \to \infty} t (\langle I_i^2 \rangle_c = \langle I_i^2(t) \rangle - \langle I_i \rangle^2)$, 
we define the squared relative uncertainty for individual currents as $\epsilon^2_i = \langle I_i^2 \rangle_c / \langle I_i\rangle^2$ and 
construct the ratio between the uncertainties of the two currents in the linear response regime, 
\be
{\cal Q} \equiv \frac{\epsilon^2_2}{\epsilon^2_1} = \frac{L_{22}}{L_{11}} \, \frac{\sum_{i j } L_{1 i} L_{1 j} \, A_{i} \, A_{j}} {\sum_{ij} L_{2 i} L_{2 j } \, A_{i} \, A_{j}}.
\label{eq:def}
\ee
Since the currents are linear in affinities, the fluctuation of currents are replaced by their 
corresponding equilibrium values in the linear response limit, \big[$\langle I^2_{i}\rangle_c\big]_{\rm eq}= L_{ii}$, ensuing from the fluctuation-dissipation relation in equilibrium.  
% \,\, \langle j^2_c \rangle_c= L_{cc}$.
%One can then write ${\cal Q}$ as 
%\be
%{\cal Q} = \frac{L_{cc}}{L_{qq}} \, \frac{\sum_{\alpha,\beta} L_{q \alpha} L_{q\beta} \, A_{\alpha} \, A_{\beta}} {\sum_{\alpha,\beta} L_{c \alpha} L_{c\beta} \, A_{\alpha} \, A_{\beta}} 
%\ee
After simple algebraic manipulations, Eq.~(\ref{eq:def}) reduces to \be
{\cal Q} =1 + \frac{1}{L_{11} \langle I_2 \rangle^2} \, \sum_{ij} \Big[ L_{1 i} L_{1 j} L_{22} - L_{2 i}  L_{2 j} L_{11} \Big] A_{i} A_{j}.
\ee
Interestingly, the above summation does not contribute for $i \neq j$ and therefore reduces to
\be
{\cal Q} =1 + \frac{1}{L_{11} \langle I_2 \rangle^2} \, \sum_{i} \Big[ L^2_{1 i} L_{22} - L^2_{2 i} \, L_{11} \Big] A^2_{i}.
\label{Eq:simplify}
\ee
The term in the sum can be written as a product of $\big(L_{22} A_2^2 - L_{11} A_1^2\big)\, \big(L_{12}^2 - L_{11} L_{22}\big)$, which in general can take an arbitrary sign. 
However, for time-reversal symmetric systems,  due to the non-negativity property of the net entropy production rate in steady state, $\langle \sigma \rangle = \sum_{ij} \langle I_i \rangle A_i = \sum_{ij} L_{ij}\, A_{i} A_{j} \geq 0$, Onsager's response coefficients satisfy the inequality $L_{12}^2 - L_{11} L_{22} \leq 0$, thus the second term in the  product above is always negative. 
So-far, we have not imposed any restriction on the operational behavior of the system. However, 
%To-realise such setup working as a useful machine, Interestingly, 
upon imposing the condition that the steady-state setup operates as an engine, it turns out the other term is also negative, $\big(L_{22} A_2^2 - L_{11} A_1^2\big)\leq 0$. 
Proof: To realize a steady-state engine, we assign $I_1$ ($I_2$) as the input (output) channel. 
Recall that, as per our convention, the current flowing into the system is considered as positive. % DDD check simulations
We therefore demand that 
% $\langle \sigma_1 \rangle \equiv  why do we need to define sigma_1?
$\langle I_1\rangle A_1 \geq 0$ which in the linear response produces the condition $L_{11} A_1^2 \geq - L_{12} A_1 A_2 $. Similarly, the system should deliver power i.e., 
%$ - \langle \sigma_2 \rangle \equiv 
$- \langle I_2 \rangle A_2 \geq 0$, which yields another condition, $-L_{12} A_{1} A_{2} \geq  L_{22} A_2^2$. Combining these two conditions we receive the inequality
\be 
L_{22} A_2^2 - L_{11} A_1^2 \leq 0,
\label{ineq:Onsager}
\ee
which holds as long as the steady-state setup delivers power. %operates as an engine. 
Back to Eq. (\ref{Eq:simplify}), we conclude that ${{\cal Q} \geq 1}$ in three operational
regimes, see Table. %for heat engines 
This is the central result of our paper, and it allows us to conclude the following:
%relationships, and bounds for $\eta^{(2)}$:

%This completes the proof for the above result.  
First, the condition ${\cal Q} \geq 1$ immediately implies that the TUR product ($\langle \sigma \rangle \frac{\langle I_i^2\rangle_c}{\langle I_i \rangle^2}$) 
for the output current is always lower-bounded by the corresponding TUR product for the input current, 
\be
\langle \sigma \rangle \frac{\langle I_2^2\rangle_c}{\langle I_2 \rangle^2}\geq \,
 \langle \sigma \rangle  \frac{\langle I_1^2\rangle_c}{\langle I_1\rangle^2}.
\label{ineq:TUR}
\ee
Earlier studies \cite{Barato:2015:UncRel,trade-off-engine} derived independent TUR bounds for the individual currents for  generic classes of steady-state systems (both Markovian and non-Markovian).
In contrast, Eq. (\ref{ineq:TUR}) shows that  %operational regime, 
these bounds are not independent in the operational regime.

Second,  an immediate consequence of  ${\cal Q} \geq 1$ is that the ratio between fluctuations of currents gets lower-bounded by the square value of the averaged efficiency,  
\be
\eta^{(2)} = \frac{ A_2^2 \langle I_2^2 \rangle_c}{A_1^2 \langle  I_1^2 \rangle_c} \geq \Big[\frac{-A_2 \, \langle I_2 \rangle}{A_1 \, \langle I_1 \rangle}\Big]^2 = \langle \eta \rangle^2,
\ee
where $\langle \eta \rangle \equiv -A_2 \langle I_2 \rangle / A_1 \langle I_1\rangle$ is the average efficiency of the machine. 
%The above inequality states that the fluctuation for a steady state machine is lower bounded by the square value of the average efficiency. 

Third, based on Eq.~(\ref{ineq:Onsager}) we  derive an upper bound for $\eta^{(2)}$ in the linear response regime. Using $L_{22}=\langle I_2^2\rangle_c$
and $L_{11}=\langle I_1^2\rangle_c$ we conclude that
$\eta^{(2)} = \frac{A_2^2 L_{22}}{A_1^2 L_{11}}  \leq 1$.
 Here, the unity on the right hand side corresponds to  the square of the scaled Carnot efficiency, thus (see Table) for an engine we write the upper bound as
 \bea
 \eta^{(2)} \leq \eta_C^2.
 \label{eq:LUB}
 \eea
 %
 % In all the above cases, %DDD?
Altogether, we find that for an engine in linear response,
$\langle\eta\rangle^2 \leq \eta^{(2)}\leq\eta_C^2$, which is our main result. 
This inequality further provides a tighter bound on the average efficiency, $\langle \eta\rangle\leq \sqrt{\eta^{(2)}}\leq \eta_C$.
This result applies for quantum or classical thermal machines, and in the Table we show
the form of the bounds for engines, refrigerators and heat pumps.
%$\langle \sigma \rangle \to 0$. 
The bounds are simultaneously saturated in the reversible limit, when the output power [Eq. (\ref{ineq:Onsager})] vanishes. % DDDD
Another interesting limit is
when the input and the output currents are tightly coupled \cite{st-thermo-Broeck}, resulting in a vanishing determinant of the Onsager response matrix, thus ${\cal Q}=1$. In this case, only the lower bound in (\ref{eq:LUB}) is saturated. Moreover, if the tight-coupling condition is satisfied for the {\it stochastic currents}, $I_1 \propto I_2$, (and not only for the averaged currents, $ \langle I_1 \rangle \propto \langle I_2\rangle$), one receives 
${\cal Q}=1$ even arbitrarily far from equilibrium, leading to $\eta^{(2)} = \langle \eta \rangle^2 \leq \eta_C^2$. Additional bounds on ratios of high order cumulants in the tight coupling limit are presented in \cite{supp}.   
The special case, when a single affinity is applied (one of the $A_i=0$) is discussed in \cite{supp}.
% DDD check conventions for A, eta
%============================================
\begin{table*}[ht]
\caption{Input ($I_1$) and output ($I_2$) currents, their affinities, average efficiency and ratio of fluctuations in different operational regimes following the notation of Ref. \cite{eff-fluc1}. The delivered power is given by $-I_2$.  $\eta_C=1-T_c/T_h$ is the Carnot efficiency.}
\centering 
\addtolength{\tabcolsep}{8pt} 
\begin{tabular}{c c c c c } 
\hline\hline
\,\,\,\,\,\,\,\,& Heat Engine \,\,\,\,\,\,\,\,\,& Refrigerator \,\,\,\,\,\,\,\,\,& Heat Pump\\ 
\hline
$ I_1$ & $  \dot{q}_h $ & $ \dot{w} $ & $ \dot{w} $ \\
$A_1$ & $\eta_C/T_c$ & $1/T_h$ & $1/T_c$ \\
$ I_2$ & $  \dot{w} $ & $ -\dot{q}_c $ & $ \dot{q}_h $ \\
$A_2$ & $1/T_c$ & $\eta_C/T_c$ & $\eta_C/T_c$ \\
$ \langle \eta \rangle$ & $\frac{-\langle \dot{w} \rangle} {\langle \dot{q}_h \rangle} \leq \eta_C$ &  $\frac{\langle \dot{q}_c \rangle}{\langle \dot{w} \rangle} \leq \frac{(1-\eta_C)}{\eta_C}$ & $\frac{\langle -\dot{q}_h \rangle}{ \langle \dot{w} \rangle} \leq \frac{1}{\eta_C}$   \\
$ \eta^{(2)} $ & $\langle \eta \rangle_{\rm eng}^2 \leq \eta^{(2)}_{\rm eng} \leq \eta_C^2$ &  $\langle \eta \rangle_{\rm ref}^2 \leq \eta^{(2)}_{\rm ref} \leq \big(\frac{1-\eta_C}{\eta_C}\big)^2 $ &  $\langle \eta \rangle_{\rm pump}^2 \leq \eta^{(2)}_{\rm pump} \leq \frac{1}{\eta_C^2}$\\
%
%${\cal Q}\geq 1$ & $\epsilon^2_{\dot{w}}/\epsilon^2_{\dot{q}_h}$ & $\epsilon^2_{\dot{q}_c}/\epsilon^2_{\dot{w}}$ &  $\epsilon^2_{\dot{q}_h}/\epsilon^2_{\dot{w}}$ \\
%$\eta^{(2)} \geq$ & $\langle 
\hline 
\end{tabular}
\label{table:affinity} 
\end{table*}
%
%For thermal machines, $\langle \eta\rangle = \langle \tilde{\eta} \rangle / \eta_c$, with $\langle \tilde{\eta} \rangle = -\langle \dot{w} \rangle / \langle \dot{q} \rangle$ being the standard definition of energy efficiency and $\eta_c$ the original Carnot efficiency. Similaly, the definition for the fluctuation for efficiency follows as $\eta^{(2)} = \frac{\tilde{\eta}^{(2)}}{\eta_c^2}$ with $\tilde{\eta}^{(2)} = \langle \dot{w}^2 \rangle_c/ \langle \dot{q}^2 \rangle_c$. Then for thermal machines the lower and upper bound implies $\tilde{\eta}^{(2)} \geq \langle \tilde{\eta} \rangle^2$ and  $\tilde{\eta}^{(2)} \leq {\eta}_C^2$.   For the efficiency one then receives $\tilde{\eta}^{(2)} = \langle \tilde{\eta} \rangle^2$. 
%
%In contrast, the upper bound for the efficiency saturates i.e., $\tilde{\eta}^{(2)}= {\eta}_C^2$  in the reversible limit occuring when $\langle \sigma \rangle \to 0$. In fact, it is easy to see that in the tight coupling limit if one can extend the above analysis trivially in this limit, in which case, for engines operating arbitrarily far-from-equilibrium, the equality will hold. 
%
%
%One can extend the above analysis can show that when the device works as a refrigerator the quantity ${\cal Q}$ is upper bounded by the value 1. For other cases, such as a ${\cal Q}$ does not follow any particular bound.
%Discussion about thermal engine:
%
% DDD What is the objective of simulations? linear reponse--> beyond. More? 
% Decide on convention for affinities and currents.
We now illustrate the bounds on $\eta^{(2)}$ for steady state autonomous thermal machines.
% within and beyond linear response. 

{\it Example I: Thermoelectric transport in two-terminal systems.-}  
We consider a two-terminal thermoelectric device depicted in Fig. \ref{fig:scheme}(a). The junction consists of an elastic scatter (e.g., array of quantum dots), which is connected to two fermionic reservoirs. 
For such a noninteracting setup the full-counting statistics for both charge and heat currents can be obtained exactly using a scattering matrix formalism. It is given by the celebrated Levitov-Lesovik formula \cite{Levitov, fcs-charge1, fcs-charge2}. Expressions of the currents and their higher order cumulants are given in \cite{supp}. 
$\mu_{L,R}$ and $T_{L,R}$ denote the electrochemical potential and temperatures of the left ($L$) and right ($R$) electronic reservoirs, $\Delta T= T_R-T_L$, with $T=(T_L+T_R)/2$ the average temperature. 
 % \cite{supp}. % DDD fixed
%$A_1= \beta (\mu_L\!-\!\mu_R) $ and $A_2= \Delta T / T^2 $, respectively, % DDD this is inconsistent with the Table. Why do we need to mention what are the A1 and A2?
To realize different operational regimes, we set the thermodynamic parameters for the reservoirs as $T_R > T_L$ and $\mu_L >\mu_R$. 
In a thermoelectric engine, the heat current absorbed from the hot (right) terminal is used to drive charge current against the chemical bias. 
%In the linear response regime, 
The thermodynamic affinities responsible for realizing an engine are %(driving charge and heat currents across the system are 
$A_1= \eta_C/T_L $ and $A_2= 1/ T_L$, see Table.
When the system operates as a refrigerator or a heat pump,
the (left) cold terminal is cooled down using electrochemical work. 
Following consistent stochastic thermodynamics framework \cite{st-thermo-Broeck}, the stochastic output power for the setup is written as $-\dot{w} \equiv (\mu_L\!-\!\mu_R)\, I^R_e$ which on average is positive (negative) for engine (refrigerator). Correspondingly, the stochastic heat current is given as $\dot{q}_{\alpha} \equiv  I^{\alpha}_q = I^{\alpha}_u - \mu_{\alpha} I^{\alpha}_e, \alpha=L,R$. Currents here are defined positive when entering the system from the right terminal. 

\begin{figure}[h]
\centering
\includegraphics[trim= 0.65cm 0.6cm 0.8cm 0.6cm, clip=true,width=\columnwidth]
{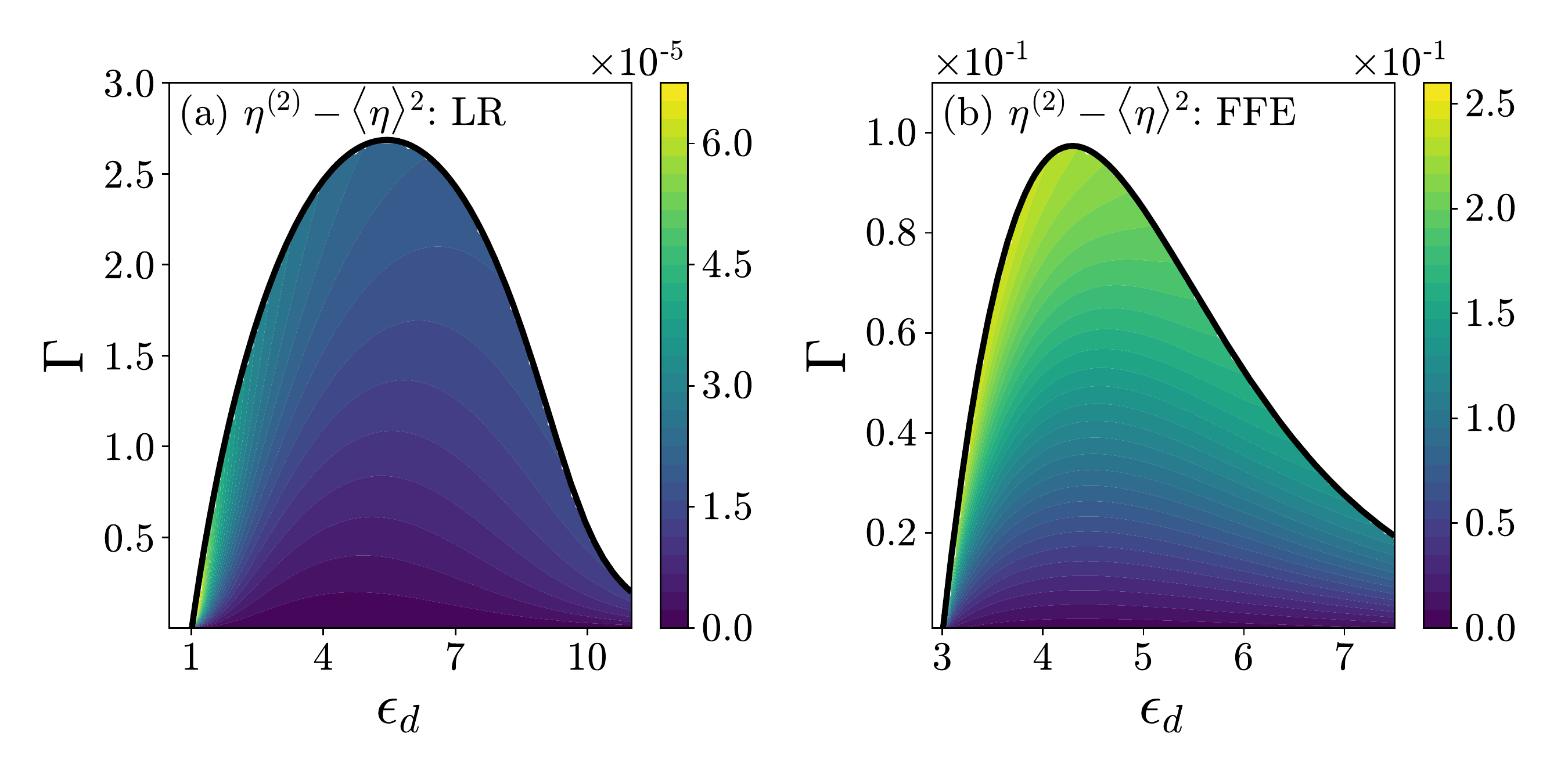}
\includegraphics[trim= 0.65cm 0.6cm 0.8cm 0.6cm, clip=true,width=\columnwidth]{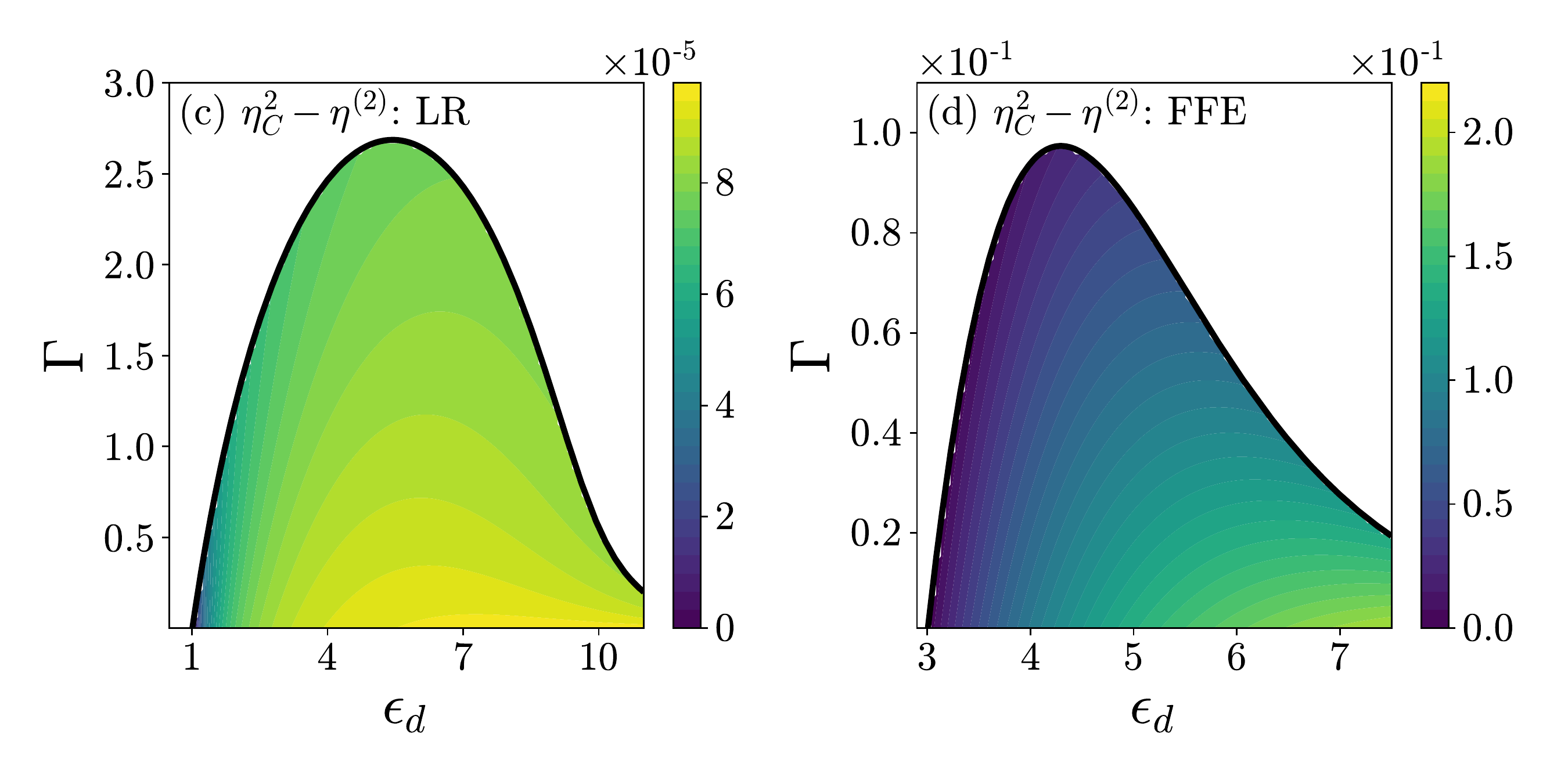}
\caption{(Color online) 
(a)-(b) Test of the lower bound, $\eta^{(2)}\geq\langle\eta\rangle^2$, and (c)-(d) the upper bound, $\eta_C^2\geq\eta^{(2)}$, 
for a single-dot thermoelectric system within linear response (LR) and far from equilibrium (FFE). We focus on the ranges of $\epsilon_d$ and $\Gamma$ at which the system operates as an engine. Positive values signify that bounds are satisfied (see Table~\ref{table:affinity}). We use $\beta_L=1.01$, $\beta_R = 1$, $\mu_L=0.01$, $\mu_R=0$ for (a) and (c) and $\beta_L = 2$, $\beta_R = 1$, $\mu_L = 1$, $\mu_R = -1$ for (b) and (d).}
\label{fig:engine-bounds}
\end{figure}

\begin{figure}[h]
\includegraphics[width=\columnwidth]{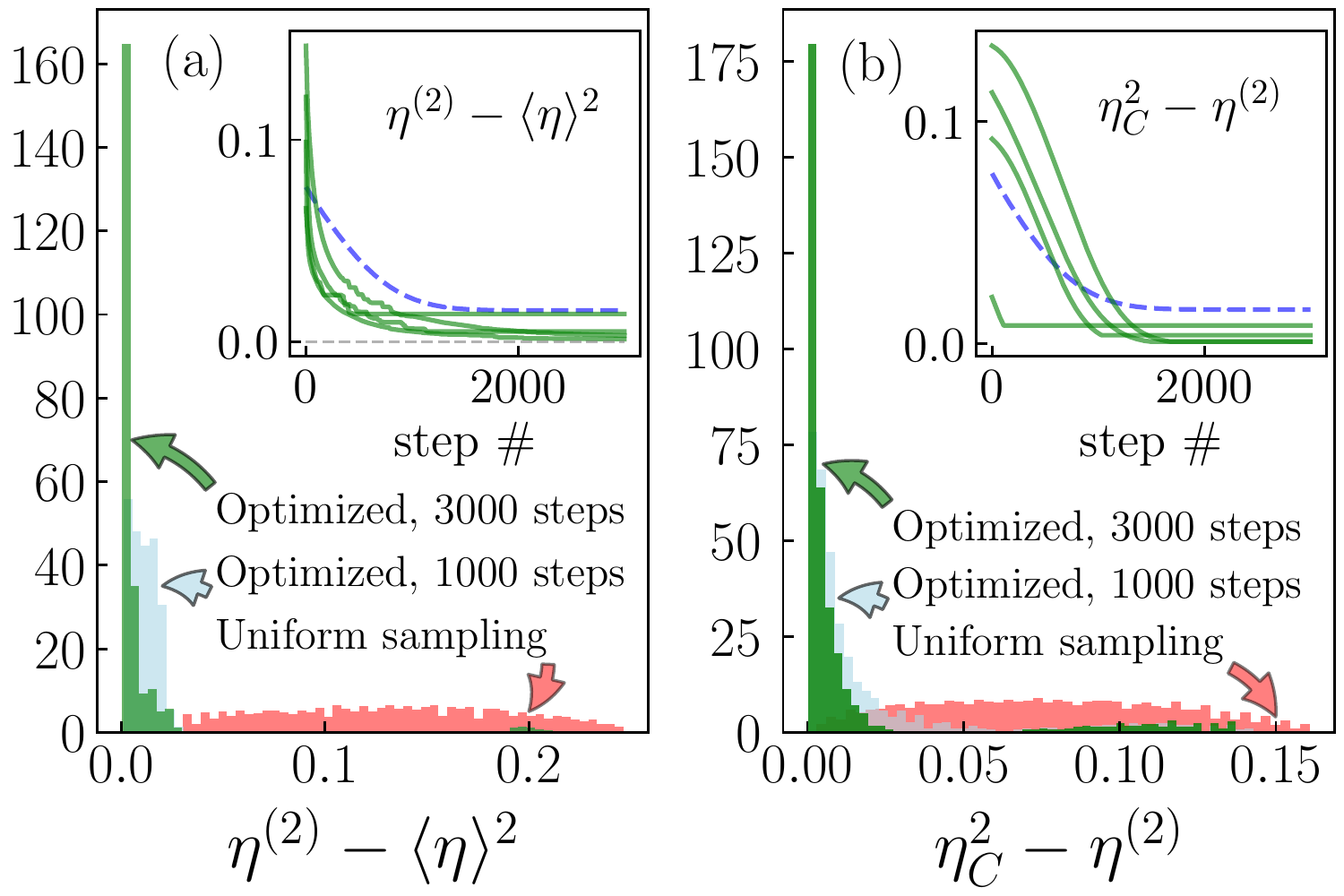}
\caption{(Color online) Distributions of the difference (a) $\eta^{(2)}-\langle \eta \rangle^2$  and (b) $\eta_C^2 - \eta^{(2)}$  for a single-dot thermoelectric junction beyond linear response. The  three cases shown are of a uniform sampling (red) of the engine region (see Fig.~\ref{fig:engine-bounds} (b) and (d)), and the result of a stochastic optimization of 1000 steps (light blue) and 3000 steps (green). We do not observe a violation of the bound throughout the minimization process. The insets show the mean value of the optimization path (in dashed blue), and samples of optimization paths for random initial points in parameter space (green). We use the same parameters as in Fig.~\ref{fig:engine-bounds}(b).}
\label{fig:eta2_optim}
\end{figure}

In Fig. \ref{fig:engine-bounds} we display contour plots of (a)-(b) $\eta^{(2)}-\langle \eta \rangle^2$ and (c)-(d) $\eta_C^2-\eta^{(2)}$ for a thermoelectric engine both in linear response (LR) and far from equilibrium (FFE). We use a single quantum dot as the scatter with the dot energy $\epsilon_d$ and the hybridization strength $\Gamma$, which is assumed to be equal for both left and right leads. These simulations exemplify the validity of the bounds. 
%%%%
To further establish the lower and upper bounds for an engine beyond the linear response regime, we employ the following numerical procedure: Beginning with a point in our parameter space ($\Gamma$ and $\epsilon_d$), 
and selecting particular biases $\Delta \mu$ and $\Delta T$, % DDDD
$\eta^{(2)}$ takes a certain value, generally away from the bounds. Focusing on the lower bound (and similarly for the upper bound), we move in parameter space according to a stochastic optimization algorithm (ADAM~\cite{Adam}),
by minimizing a cost function, which is a function of $\eta^{(2)}-\langle \eta \rangle^2$ (see \cite{supp} for details). This process is performed for a large ensemble of around 1100 initial conditions, near and inside the engine domain.
In Fig.~\ref{fig:eta2_optim}, we show the resulting distribution of the minima $\eta^{(2)}-\langle \eta \rangle^2$ and $\eta_C^2 - \eta^{(2)}$. We begin with a uniform sample of points inside the engine regime. We then present histograms after 1000 and 3000 optimization steps. As the number of steps increases, more points are nearing saturation of the bound as expected for the optimization process.
In the insets we illustrate that for each step the differences above are positive throughout the optimization process by plotting four random minimization paths, along with the average of all paths.
This powerful procedure provides numerical evidence for the validity of the bounds for the single dot engine beyond linear response. 
% still one voltage and delta T... 
Furthermore, it reveals that the lower bound is saturated in the tight coupling limit, as values for $\eta^{(2)}-\langle \eta \rangle^2$ after minimization 
correspond to diminishing $\Gamma$.

%These simulations suggest that the lower bound for $\eta^{(2)} $ is satisfied not only in linear response, but {\bf even beyond}.
%%%%%
We complement these results in \cite{supp} with (i)  simulations demonstrating that analogous  bounds for $\eta^{(2)}$ are satisfied for single-dot refrigerators and pumps in linear response, along with examples suggesting the validity of the bounds in the FFE regime.
% (ii) {\textcolor{red}{similar simulations for heat pumps.}}
(ii) Simulations of thermoelectric engines based on a serial double-dot system. Here, we show that the lower and upper bounds for $\eta^{(2)}$ hold beyond linear response, even when the standard Markovian TUR ($\langle \sigma \rangle \frac{\langle I_i^2\rangle_c}{\langle I_i \rangle^2}\geq 2$) is violated.

%=======================
%==============================================
% Figure 2
% Matthew: Can you add text 
%(a) engine; linear response 
%(b) engine; far-from-equilibirum
% (c) refrigerator; linear response 
%(d) refrigeration; far from eq.
%=============================================================
% Figure 3
%======================================================================
% Figure 4
\begin{figure}[h]
\centering
{\includegraphics[width=2.60cm] {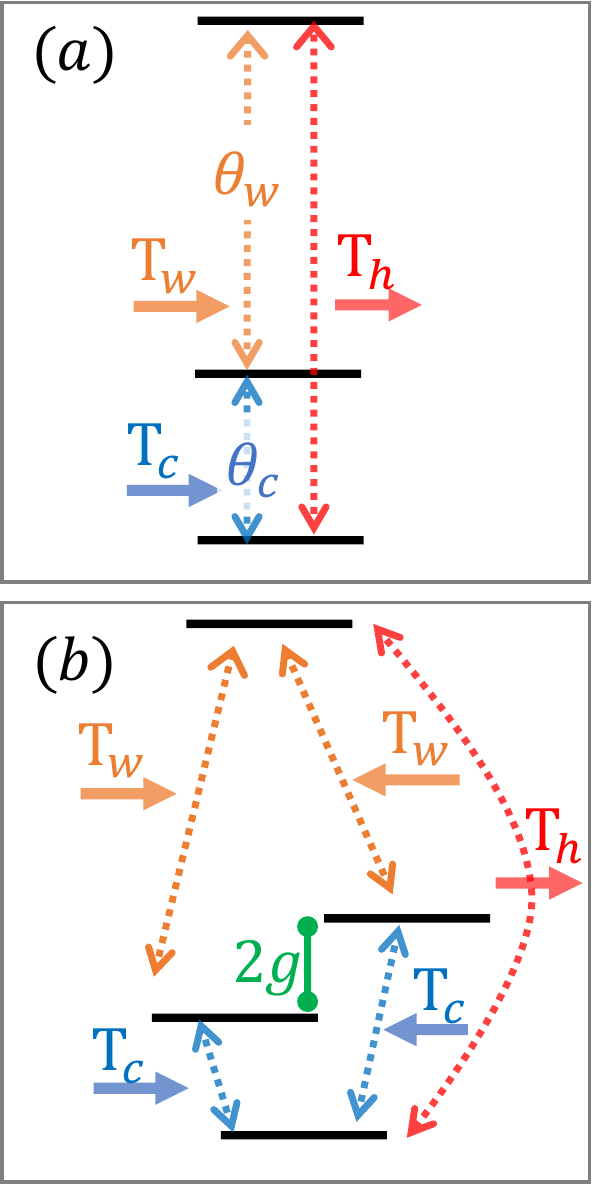} %diag-newN3-crop.pdf} 
\hspace{2mm}
\includegraphics[width=5.6cm]{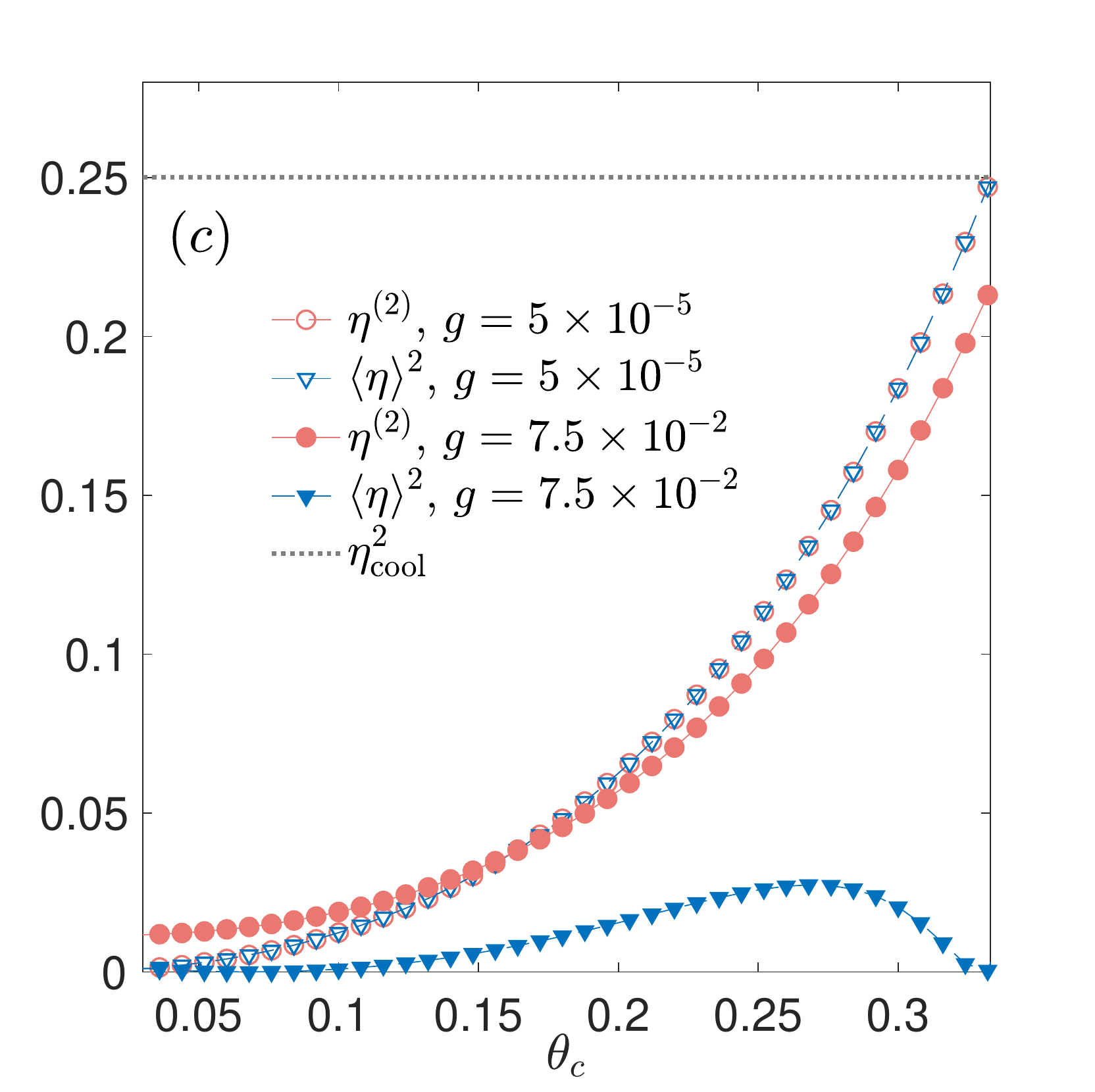}}
\caption{(Color online) %{\textbf{ We don't define $\eta_{cool}$ here, figure has $\theta_c$, caption has $\theta_C$}...}
(a) Three-level and (b) four-level models for QARs. Dashed arrows represent bath-induced transitions. Full-line arrows show the direction of heat absorbed and released when the system operates as a QAR.
(c) Simulations of $\eta^{(2)}$ for the four-level model (b). Here, $\theta_c$ represents the energy gap between the ground level to halfway between the intermediate levels. We use $T_w=0.4$, $T_h=0.2$, $T_c=0.1$, with the maximum cooling efficiency  $\eta_{\rm cool}\equiv\frac{\beta_h-\beta_w}{\beta_c-\beta_h}=0.5$. The spectral function of the baths are Ohmic, $J_b(\omega)=\gamma_b \omega e^{-|\omega|/\Lambda}$, with $\gamma_{h}=2\times 10^{-3}$, $\gamma_{w,c}=10^{-3}$, $\Lambda=50$.
}
\label{fig:QAR}
\end{figure}
%=======================
{\it Example II: Quantum Absorption Refrigerators.}
% Intro
Models for QARs have served a crucial role in establishing working principles of autonomous quantum thermal machines \cite{Kosloff, Luis, MarkRev14} with recent experiments realizing such models using trapped ions \cite{engine-refg,ionE}. %and circuit QED \cite{Pekola}.
In QARs, heat is extracted from a cold ($c$) bath
and released into a hot ($h$) environment by utilizing heat from a so-called work ($w$) bath. The reversed operation realizes a heat engine. 
We identify three temperatures, $T_w>T_h>T_c$ in a QAR, and three heat currents, $I_{c,h,w}$, defined positive when flowing towards the system.
%with two affinities, $\beta_h-\beta_c$ and $\beta_h-\beta_w$; $\beta=1/T$. 
Schematic diagrams of three-level and four-level QARs are displayed in Figs. \ref{fig:QAR}(a) and \ref{fig:QAR}(b), respectively.
% Here
In what follows, we limit our discussion to the weak system-bath coupling limit, which can be handled with perturbative quantum Master equations.
While so-far most studies characterized QARs based on their averaged currents \cite{KilgourQAR}, fluctuations of currents in QARs were analyzed recently in Refs. \cite{Segal18,Junjie21}, demonstrating e.g. the nontrivial impact of quantum coherences on the cooling current and its fluctuation.

%defintions
The cooling efficiency of QARs is defined as $\langle \eta\rangle = \langle I_c\rangle/\langle I_w\rangle$. It is bounded by $\langle \eta\rangle \leq\eta_{\rm cool}$
where $\eta_{\rm cool}\equiv\frac{(\beta_h-\beta_w)}{\beta_c-\beta_h} \xrightarrow[]{T_w\gg T_h} (1-\eta_C)/\eta_C$ \cite{Luis}.
%When $T_w\gg T_h$ we recover the Carnot bound for a refrigerator,
%$\eta_{cool}=(1-\eta_C)/\eta_C$.
%
% 3LQAR
For a three-level QAR with energy levels sketched in Fig. \ref{fig:QAR}(a), we get  $\langle \eta\rangle=\theta_c/\theta_w$.
Here, $\theta_{c,w}$ are energy gaps with transitions activated by the cold and work baths, respectively.
Furthermore, based on the formalism of the truncated cumulant generating function  \cite{Segal18}, we obtain here the fluctuations of the currents, 
%\bea
$\eta^{(2)}\equiv \frac{\langle I_c^2\rangle_c}{\langle I_w^2\rangle_c} = \left(\frac{\theta_c}{\theta_w}\right)^2$.
%\eea
%
We therefore conclude that the three-level QAR  
%and in linear response  
satisfies at weak coupling
$\langle\eta\rangle^2=\eta^{(2)}\leq\eta_{\rm cool}^2$.
The upper limit is deduced based on our analysis of linear response.
The saturation of the lower bound holds even in the far-from-equilibrium regime, and it corresponds to the tight-coupling limit, which holds under the weak coupling approximation.

To explore behavior beyond the tight-coupling limit, we sketch in Fig. \ref{fig:QAR}(b) a four-level QAR with non-degenerate intermediate levels. At weak coupling, the two cooling cycles compete. As a result, the maximal efficiency $\eta_{\rm cool}$ cannot be achieved and the system always dissipates heat \cite{Correa15}. We simulate fluctuations of the currents in the incoherent limit \cite{Segal18,supp},
%by adopting the computational procedure introduced in Refs. \cite{Segal18,Junjie21}.
%For simplicity, we simulate the system in the incoherent limit \cite{supp}.
and exemplify the behavior of this model in Fig. \ref{fig:QAR}(c). As expected, 
deviations from the tight-coupling limit become more pronounced
as we increase the gap between the intermediate levels.
The extreme case $g\ll \gamma\theta_{c,w}$ is presented to illustrate the saturation of the bound in the tight coupling limit, $\eta^{(2)}=\langle\eta\rangle^2$, further achieving the upper bound $\eta_{\rm cool}^2$ at a certain value.
As we increase the gap, $g$, the two cycles compete, the tight-coupling limit no longer holds, and we observe nontrivial results: Both lower and upper bounds
for $\eta^{(2)}$ are satisfied---even beyond the linear response regime.  
We further confirmed with simulations the validity of these bounds in broad range of parameters $g$, $\gamma$ and $\theta_{c,w}$. 
%In \cite{supp} we include explanations over the method and additional simulations.

%\vspace{5cm}

{\it Perspective.--} 
Beyond the averaged behavior of observables, the performance of small thermal machines should be characterized  by their fluctuations.
In this work, we proved that in linear response, relative fluctuations of the output current (power in heat engines) are greater than fluctuations in the input current (absorbed heat) in the thermal engine regime. We immediately translated this inequality into universal
lower and upper bounds on $\eta^{(2)}$, the ratio between the fluctuations of output and input channels, which hold in linear response thermodynamics. Numerical simulations of quantum dots thermoelectric junctions and quantum absorption refrigerators suggest the broader validity of these bounds. 
%
%Future work will be focused on 
Proofs of bounds on fluctuations and higher order cumulants beyond linear response for general (possibly quantum) dynamics
%derivations of bounds for higher-order cumulants,
%and the generalisation of this study for time-reversal breaking systems 
remain as an open challenge.

%===================================================================================================
BKA acknowledges the MATRICS grant MTR/2020/000472 from SERB, Government of India. BKA and DS thank the Shastri Indo-Canadian Institute for providing financial support for this research work in the form of a Shastri Institutional Collaborative Research Grant (SICRG).
DS acknowledges support from an NSERC Discovery Grant and the Canada Research Chair program.
The work of IK was supported by the CQIQC at the University of Toronto.
The research of MG was funded by an NSERC Graduate Scholarship - Master's (CGS M).
SS acknowledge support from the Council of Scientific \& Industrial Research (CSIR), India (Grant Number 1061651988).  

\include{supp}
\end{document}

%% file: supp.tex
\title{Supplemental Material: Universal Bounds on Fluctuations in Continuous Thermal Machines}

\author{Sushant Saryal}
\affiliation{Department of Physics,
		Indian Institute of Science Education and Research, Pune 411008, India}
\author{Matthew Gerry}
\affiliation{Department of Physics, University of Toronto, Toronto, Ontario, Canada M5S 1A7}	
\author{Ilia Khait}
\affiliation{Department of Physics, University of Toronto, Toronto, Ontario, Canada M5S 1A7}		
	\author{Dvira Segal}
\affiliation{Department of Chemistry and Centre for Quantum Information and Quantum Control,
University of Toronto, 80 Saint George St., Toronto, Ontario, M5S 3H6, Canada}
\affiliation{Department of Physics, University of Toronto, Toronto, Ontario, Canada M5S 1A7}
		\email{dvira.segal@utoronto.ca}
\author {Bijay Kumar Agarwalla}
\affiliation{Department of Physics,
		Indian Institute of Science Education and Research, Pune 411008, India}
		\email{bijay@iiserpune.ac.in}

%\email{dvira.segal@utoronto.ca}
%==============================================
\date{\today}
\maketitle
\end{comment}
\begin{widetext}
\vspace{5mm}
\renewcommand{\theequation}{S\arabic{equation}}
\renewcommand{\thefigure}{S\arabic{figure}}
\renewcommand{\thesection}{S\arabic{section}}

\setcounter{equation}{0}  % reset counter
\setcounter{figure}{0}

\begin{center} 
{\Large Supplemental Material: Universal Bounds on Fluctuations in Continuous Thermal Machines}
\end{center}
%=================================================================

\section{Thermoelectric transport in noninteracting quantum junctions}

\subsection{Expressions for currents and noise from full counting statistics}

In this section, we provide expressions for cumulants 
or charge and energy currents for generic quantum noninteracting thermoelectric junction. This setup is considered in the main text to illustrate our main findings. The exact cumulant generating function (CGF) for this problem was obtained before with a scattering matrix approach. It is given by the celebrated Levitov-Lesovik formula \cite{Levitov, fcs-charge1, fcs-charge2}. The CGF is defined from the characteristic function ${\cal Z}(\chi_e, \chi_u)= \langle \exp \big(i \chi_e n_e + i \chi_u h_u) \rangle$, where $\chi_{e}$ and $\chi_u$ are the counting fields that keep track of net charge $n_e$ and energy transfer $h_u$ processes, respectively, in the right terminal. In the long-time limit, the CGF scales extensively with the integrated time $t$. The corresponding (scaled) cumulant generating function (CGF), i.e., $ G(\chi_e, \chi_u)= \lim_{t \to \infty} \frac{1}{t} \ln Z(\chi_e, \chi_u)$, is given as,
\bea
&& G(\chi_e, \chi_u) \!=\! \int_{-\infty} ^{\infty}\!\!
\frac{d\epsilon}{2\pi} \,\ln \Big( 1\!+\!{\mathcal T}(\epsilon)
\Big\{ f_R(\epsilon)[1\!-\! f_L(\epsilon)] \nonumber\\
&& \,[e^{i(\chi_e+\epsilon\chi_u)}\!-\!1] + f_L(\epsilon)[1\!-\!f_R(\epsilon)][e^{-i(\chi_e+\epsilon\chi_u)}\!-\!1]
\Big\}
\Big).
\eea 
% DDD: Can we use "characteristic function for Z, and CGF for G? We do not need then the "scaled CGF".
%
%where $\chi_{e}$ and $\chi_u$ are respectively the counting fields that keep track of charge and energy transfer processes in the right reservoir. 
As per our convention, currents flowing out of the right reservoir is considered positive.  Here, $\mathcal T(\epsilon)$ is the transmission function, which indicates the probability for electrons to transfer from the right to the left reservoir via the elastic scattering region. $f_{\alpha}(\epsilon)=[e^{\beta_{\alpha}(\epsilon-\mu_{\alpha})} +1 ]^{-1}$, $\alpha=L,R$ is the equilibrium Fermi-Dirac distribution function for the $\alpha$-th reservoir. 
From the CGF, we obtain cumulants of currents by taking derivatives with respect to the corresponding counting fields, and obtain
\begin{equation}
\label{eq:jj}
\langle I_{K}\rangle~=~\int_{-\infty}^{\infty}\frac{d\epsilon}{2\pi}\xi_{K} \, {\cal T}(\epsilon) \, \big[f_R(\epsilon) - f_L(\epsilon)\big],
\end{equation}
where $\xi_{K}=1$ $(\epsilon)$ for $K=e$ $(u)$. 
Fluctuations i.e., the second order cumulants are given by,
\begin{eqnarray}
\label{eq:noise}
\langle I^2_{K}\rangle_c 
%&\equiv& \lim_{t \to \infty} \frac{1}{t} \langle n_K^{2} \rangle_c
&=& \int_{-\infty}^{\infty}
\frac{d\epsilon}{2\pi}\xi_{\alpha}^2 \,  \mathcal{T}(\epsilon)\{f_L(\epsilon)\left[1-f_R(\epsilon)\right] \nonumber\\
&&+f_R(\epsilon)\left[1\!-\!f_L(\epsilon)\right]\}\!-\!\mathcal{T}^2(\epsilon)\left[f_R(\epsilon)\!-\!f_L(\epsilon)\right]^2.
\label{eq:second-cum}
\end{eqnarray}
Note that the heat current ($q$) cumulants can be simply obtained by setting $\xi_{q} = \epsilon-\mu_R$. In the linear response regime, one considers the limit $|\Delta T|  \ll T$ and $|\Delta \mu| \ll T$. Here, $\Delta \mu = \mu_L-\mu_R$, $\Delta T= T_R-T_L$, and $T=(T_L+T_R)/2$, $\mu=(\mu_L+\mu_R)/2$ being the average temperature and average chemical potential, respectively. The currents can then be expressed in terms of Onsager's transport coefficients,
\bea
\langle I_{e}\rangle &=& L_{ee} \left(\frac{-\Delta \mu}{T}\right) +  L_{eq} \left(\frac{\Delta T}{T^2}\right) \nonumber \\
\langle I_{q}\rangle &=& L_{qe} \left(\frac{-\Delta \mu}{T}\right) +  L_{qq} \left(\frac{\Delta T}{T^2}\right),
\eea
where we used 
$-\frac{\partial f}{\partial \epsilon}=f(1-f)\beta$. Here, $f(\epsilon)$ is the equilibrium distribution function evaluated at $T$ and $\mu$. 
The various transport coefficients are given by 
\bea
L_{ee} &=& \int_{-\infty}^{\infty} \frac{d\epsilon}{2\pi}\, \mathcal{T}(\epsilon)\, f(\epsilon) \, [1-f (\epsilon)] \nonumber \\
L_{eq} &=& L_{qe} =  \int_{-\infty}^{\infty} \frac{d\epsilon}{2\pi}\, \big(\epsilon-\mu)\,  \mathcal{T}(\epsilon)\, \, f(\epsilon) \, [1-f (\epsilon)] \nonumber \\
L_{qq} &=&\int_{-\infty}^{\infty} \frac{d\epsilon}{2\pi}\, \big(\epsilon-\mu)^2\, \mathcal{T}(\epsilon)\,   f(\epsilon) \, [1-f (\epsilon)] \nonumber \\
\eea
In the equilibrium limit ($ \Delta \mu \!=\! \Delta T \!=\!0 $), one recovers the standard equilibrium fluctuation-dissipation relation from  Eq.~(\ref{eq:second-cum}),
\bea
\Big[\langle I_q^2 \rangle_{c}\Big]_{\rm eq} = L_{qq}, \quad \Big[\langle I_e^2 \rangle_{c}\Big]_{\rm eq} = L_{ee}.
\eea
We now discuss three operational regimes for such a setup by setting $T_R > T_L$ and $\mu_L > \mu_R$. 
%==============================================
%Figure S1
%refrigerator
\begin{figure}[t]
\centering
\includegraphics[trim= 0 0.5cm 0 0, width=0.65\textwidth]
%[trim= 0.65cm 0.6cm 0.8cm 0.6cm, clip=true,width=\columnwidth]
{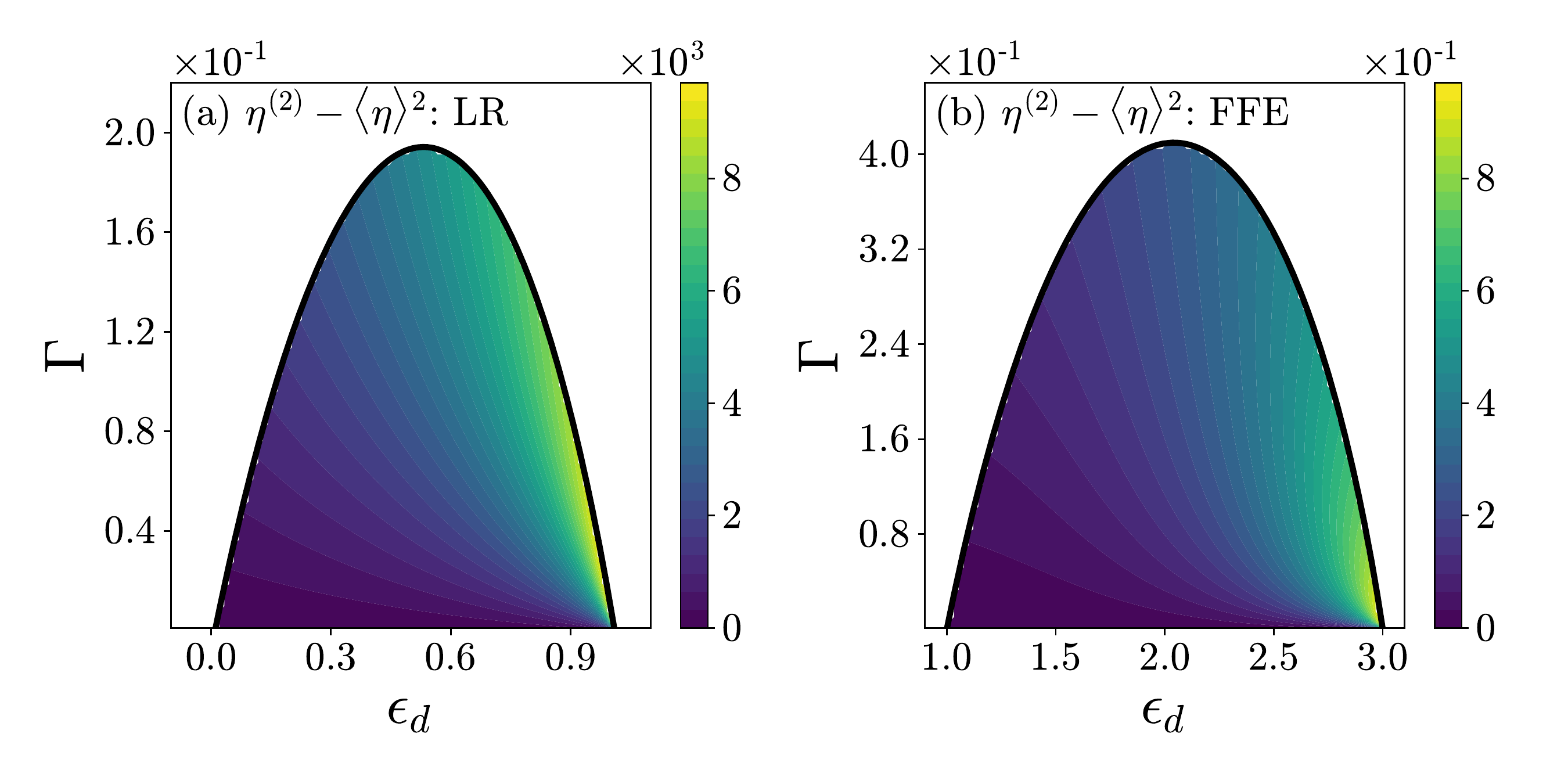}
\includegraphics[trim = 0 0 0 0.5cm, width=0.65\textwidth]
%[trim= 0.65cm 0.6cm 0.8cm 0.6cm, clip=true,width=\columnwidth]
{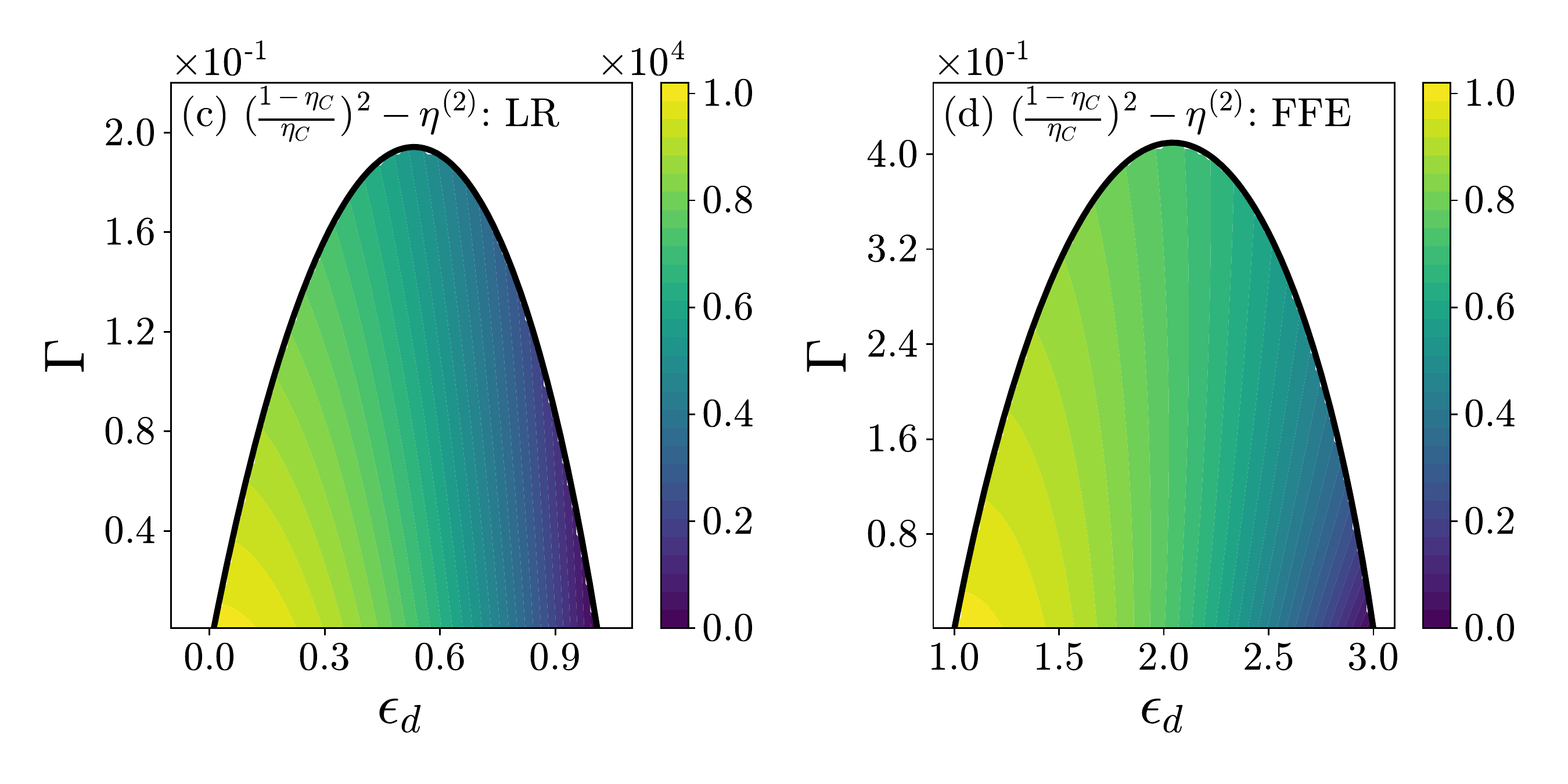}
\caption{(Color online) Single-dot thermoelectric refrigerator. 
(a)-(b) Test of the lower bound, $\eta^{(2)}\geq\langle\eta\rangle^2$ within linear response and far from equilibrium. 
(c)-(d) Test of the upper bound, $\left(\frac{1-\eta_C}{\eta_C}\right)^2\geq\eta^{(2)}$, for the same system. 
We focus on relevant ranges of $\epsilon_d$ and $\Gamma$ at which the system operates as a refrigerator. 
Positive values signify that the bound is satisfied (see Table in main text). We used $\beta_L=1.01$, $\beta_R = 1$, $\mu_L=0.01$, $\mu_R=0$ for (a) and (c) and $\beta_L = 2$, $\beta_R = 1$, $\mu_L = 1$, $\mu_R = -1$ for (b) and (d).}
\label{ref-bounds}
\end{figure}
 
%====================================================
% Figure S2
% Heat pump
\begin{figure}[h]
    \centering
      \includegraphics[trim = 0 0.5cm 0 0, width=0.65\textwidth]{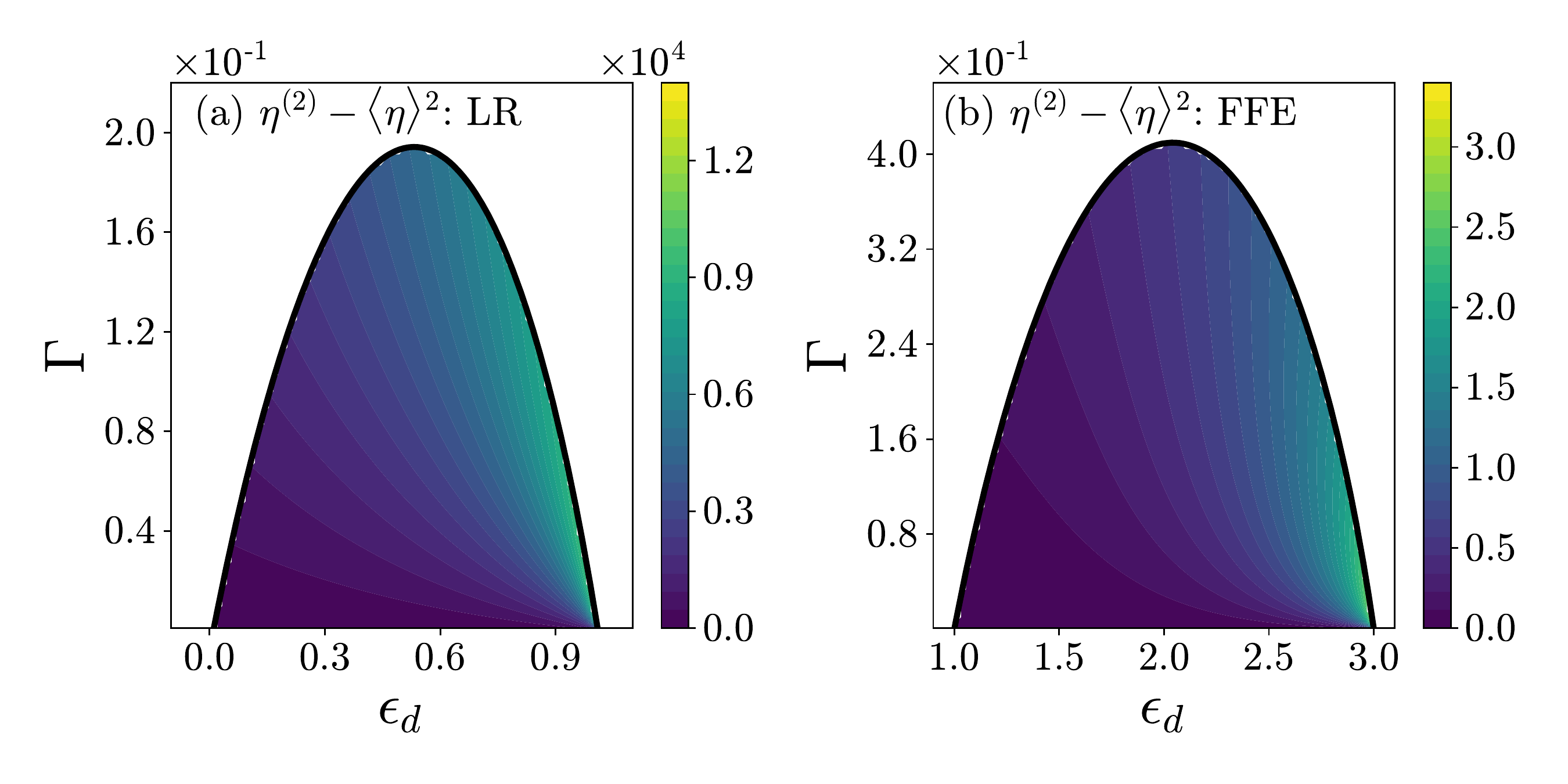}
   \includegraphics[trim = 0 0 0 0.5cm, width=0.65\textwidth]{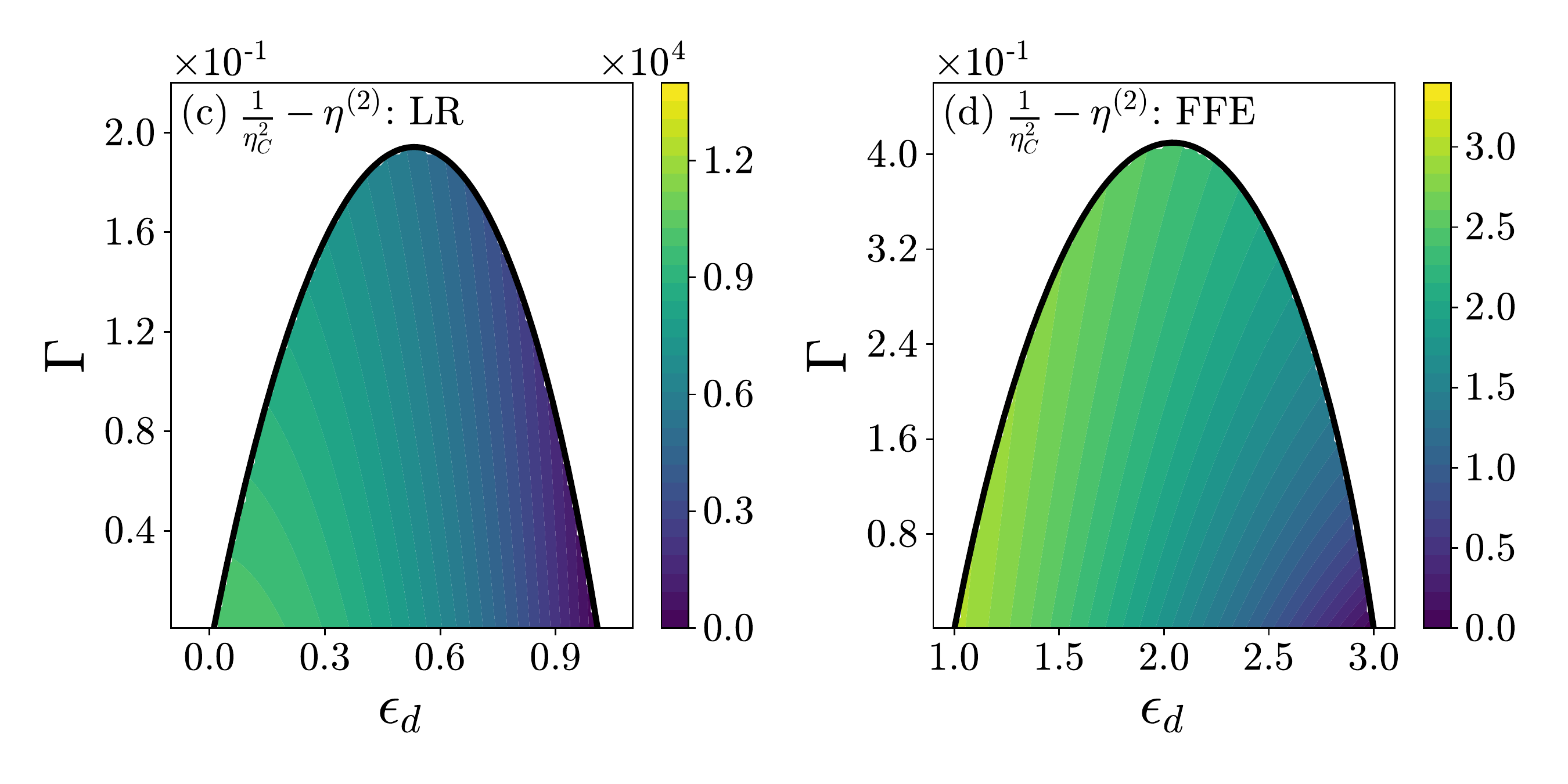}
    \caption{(Color online)
    Single-dot thermoelectric heat pump. 
    (a)-(b) Test of the lower bound, $\eta^{(2)}\geq\langle\eta\rangle^2$ within linear response and far from equilibrium. 
    (c)-(d) Test of the upper bound, $\frac{1}{\eta_C^2}\geq\eta^{(2)}$, for the same system. 
    We focus on relevant ranges of $\epsilon_d$ and $\Gamma$ at which the system operates as a heat pump. 
    Positive values signify that the bound is satisfied (see Table in main text). We used $\beta_L=1.01$, $\beta_R = 1$, $\mu_L=0.01$, $\mu_R=0$ for (a) and (c) and $\beta_L = 2$, $\beta_R = 1$, $\mu_L = 1$, $\mu_R = -1$ for (b) and (d).}
    \label{fig:pump_bounds}
\end{figure}

% Figure S3
% Q
\begin{center}
\begin{figure}[h]
 \includegraphics[width=0.75\textwidth]{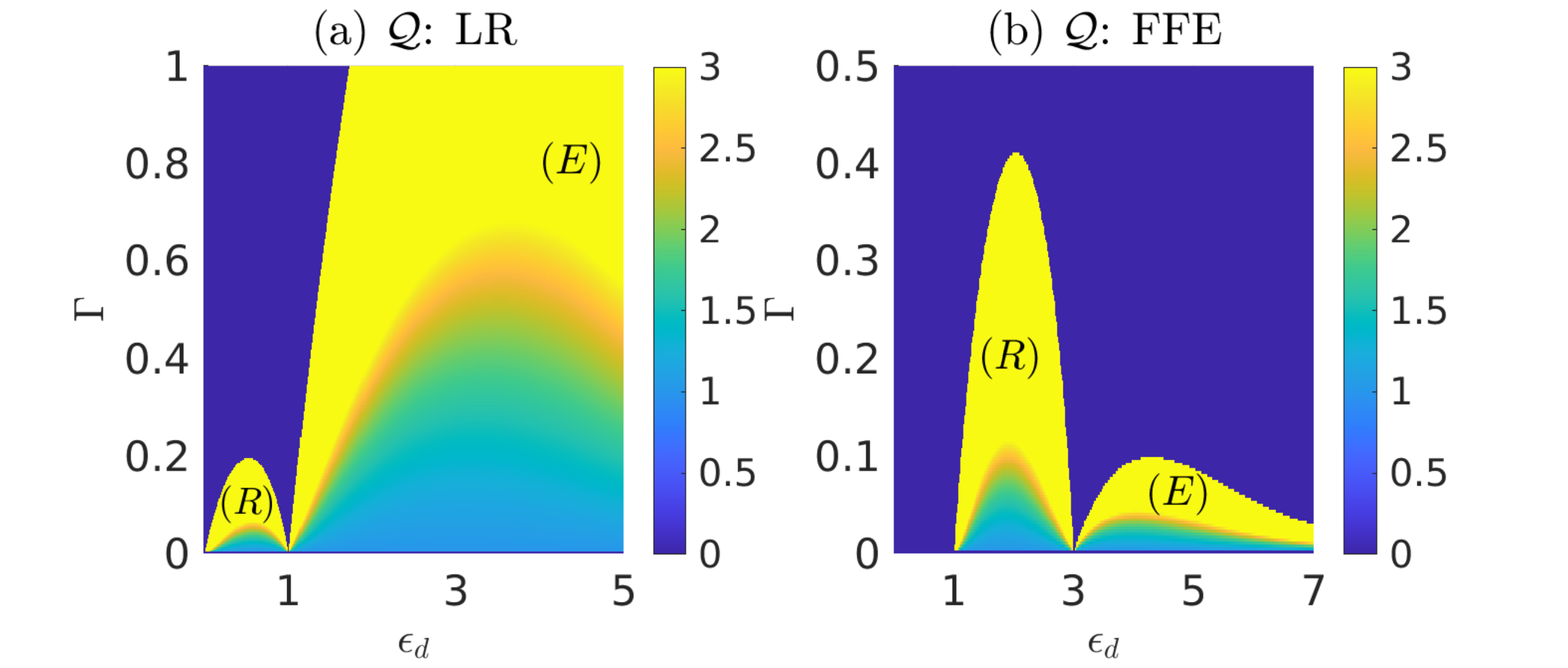}
\caption{(Color online) Plot of the ratio ${\cal Q}$ for a single dot thermoelectric system within and beyond linear response. We display both the engine (E) and refrigerator (R) regimes for the relevant ranges of $\epsilon_d$ and $\Gamma$. (a) Linear response simulations with $\beta_L=1.01$, $\beta_R=1$, $\mu_L=0.01$, $\mu_R=0$. (b) Beyond linear response simulations with  $\beta_L=2$, $\beta_R=1$, $\mu_L=1$, $\mu_R=-1$. We confirm that in both the engine and refrigerator regimes, ${\cal Q} \geq 1$. For clarity of presentation, values of ${\cal Q} > 3$ were assigned the value 3.}
\label{fig:Q-thermo}
\end{figure}
\end{center}
%==============================================================

\vspace{0.5cm}
1. {\it Thermoelectric engine}:  In order to operate the setup as a thermoelectric engine, one must ensures that the net heat current absorbed from the hot  reservoir $(R)$ is positive i.e.,  $  \langle I_u \rangle -\mu_R \langle  I _e \rangle \geq 0$. The heat absorbed  leads to transport of charge current against the chemical bias, thus delivering power i.e., $-\langle \dot{w} \rangle =(\mu_L \!-\! \mu_R) \langle I_e \rangle \geq 0$. The corresponding efficiency is then $\langle \eta\rangle_{\rm eng} = \frac{(\mu_L \!-\! \mu_R) \langle I_e \rangle}
{ \langle I_u \rangle -\mu_R \langle  I _e \rangle}$.
%-\langle \dot{w} \rangle/
%\langle \dot{q_h} \rangle$.

%Currents flowing into the system from the $R$ side is considered positive. 
\vspace{0.5cm}
2. {\it Refrigerator:}  In the refrigerator regime, heat current from the cold (left) reservoir is extracted by using the charge current flowing from high to low bias i.e., $\langle I_u \rangle -\mu_L \langle  I _e \rangle \leq 0$ and
$\langle I_e \rangle\leq0$.  The efficiency is given as  
$\langle \eta\rangle_{\rm ref} = \frac{-(\langle I_u \rangle -\mu_L \langle  I _e) \rangle }{-(\mu_L-\mu_R)\langle I_e \rangle}$.

\vspace{0.5cm}
3. {\it Heat pump:} To realize a heat pump, we demand that the heat current  
should flow towards the (right) hot reservoir, $\langle I_u \rangle -\mu_R \langle  I _e \rangle \leq 0$, by using the charge current,
$\langle I_e \rangle\leq0$. The corresponding efficiency is 
$\langle \eta\rangle_{\rm pump} = \frac{
-(\langle I_u \rangle -\mu_R \langle  I _e \rangle)}
{-(\mu_L-\mu_R)\langle I_e\rangle}$.
\vspace{0.5cm}
%XXXX--------------------------

In the main text, we illustrate the validity of the lower and upper bounds for $\eta^{(2)}$ for a thermoelectric engine, within the linear response (LR) regime and far from equilibrium (FFE). Here, in Fig.~\ref{ref-bounds} we  display the validity for the corresponding lower, $\langle\eta\rangle^2\leq\eta^{(2)}$, and upper, $\eta^{(2)}\leq(\frac{1-\eta_C}{\eta_C})^2$, bounds for a single-dot thermoelectric system in the refrigerator regime. Figure \ref{fig:pump_bounds}
presents analogous results for single-dot heat pumps.

In Fig.~\ref{fig:Q-thermo}, we show the ${\cal Q}$ ratio in the same regime as displayed in Figs.~\ref{ref-bounds}-\ref{fig:pump_bounds} both in the linear and beyond linear response regime. We observe the existence of the lower bound i.e., ${\cal Q}\geq 1$ in both the engine (E) and the refrigerator (R) regimes. %Note that, the ${\cal Q}$ values greater than 3 presented by the same  the  %In Fig.~() and () we show the corresponding lower bounds for the engine and the refrigerator. Interestingly, the lower bound is also found to be valid beyond linear response. 

\subsection{Tight-coupling limit beyond linear response}
%We also need the steady-state average entropy production rate for the engine, given by
%\bea
%\langle \sigma \rangle = \langle j_e \rangle \left(\beta_R \, \mu_R- \beta_L \, \mu_L \right)  + \langle j_u \rangle \left(\beta_L - \beta_R \right)\geq 0.
%\eea
In this subsection we focus on
the tight coupling (TC) limit. We prove the upper bound $\eta^{(2)}\leq \eta_C^2 $ and that the lower bound is saturated. Both results hold beyond linear response. We further derive bounds on ratios (output power to heat) of high order cumulants.

In the tight-coupling limit, the energy current flowing through the junction is proportional to the electron flux. While commonly, this limit is assumed for the averaged currents, here we enforce it for the stochastic currents. i.e., $ I_u = \epsilon_d \, I_e$ with $\epsilon_d$  the  energy of the single quantum dot. Since the currents here are proportional (coupled), this limit is commonly referred to as the  ``tight-coupling limit".  Such a situation can be easily realized when a single quantum level is weakly coupled to the reservoirs. The dynamics and thermodynamics of the model can be obtained from stochastic thermodynamics using Markovian master equations. 
The TC limit is of great interest in nanoelectronics since the system in this limit has the ability to operate at the Carnot efficiency \cite{Linke}. Furthermore, various universal features in nonlinear response such as the universality of the efficiency at maximum power were proved in this regime \cite{MPE}.

We  now study the ratio between work and heat fluctuations in the TC limit for a thermoelectric engine, arbitrarily far from equilibrium. First, we calculate the average thermodynamic efficiency in the TC limit, 
\bea
\langle \eta\rangle &=& \frac{-\langle \dot w\rangle}{\langle \dot q_h\rangle } 
\nonumber\\
&=& \frac{\Delta \mu \langle I_e\rangle }{\langle I_u\rangle-\mu_R \langle I_e\rangle}=
\frac{\Delta \mu}{(\epsilon_d-\mu_R)}. 
\eea
The upper bound for the mean efficiency is given by the Carnot value $\eta_C = 1 - \frac{T_L}{T_R}$. This can be  proved by enforcing the positivity requirement for the average entropy production rate. In steady state, it is defined as $\langle \sigma \rangle = \sum_{i = e, u} {A}_i \langle I_i \rangle$ where $A_e =  \left(\beta_R \, \mu_R- \beta_L \, \mu_L \right)$ and $A_u =  \left(\beta_L - \beta_R \right)$ are the two thermodynamic affinities corresponding to particle and energy current, respectively \cite{comment}.  
In the TC limit, this reduces to 
\bea
\langle \sigma\rangle =  \big[ \beta_L (\epsilon_d-\mu_L) -\beta_R (\epsilon_d-\mu_R)\big]\langle I_e\rangle \geq 0,
\eea
which is always non-negative. The zero net entropy production limit corresponds to a quasi-static situation, which, interestingly, in TC limit can be achieved without requiring the individual affinities to go to zero. The above condition in the context of thermoelectric engine yields,
%At equilibrium, both charge and energy currents vanish. At this point, $\langle \sigma\rangle=0$, or equivalently $f_L(\epsilon_d)=f_R(\epsilon_d)$, where the  Fermi functions  $f_{K}(\epsilon)=[e^{\beta_K(\epsilon-\mu_K)} +1 ]^{-1}$ ($K=L,R$), are evaluated at the energy of the dot,This translates to
%
\bea
\beta_L(\epsilon_d-\mu_L) \geq \beta_R (\epsilon_d-\mu_R),
\eea
which implies that 
\bea
\langle \eta\rangle = \frac{\Delta \mu}{(\epsilon_d-\mu_R)} \leq  1-\frac{\beta_R}{\beta_L}= \eta_C.
\eea
%Reorganizing it we identify the parameters leading to the Carnot efficiency,
%\bea
%\eta_C\equiv 1-\frac{\beta_R}{\beta_L} = \frac{(\mu_L-\mu_R)}{\epsilon_d-%\mu_R}.
%\eea
%
As expected, the mean efficiency of the engine is limited by the Carnot value $\langle \eta\rangle \leq \eta_C$.

%We now continue with the fluctuations of the power and heat currents,
%\bea
%\langle j_p^2 \rangle - \langle j_p\rangle ^2 
%=(\Delta \mu)^2 \left( \langle j_c^2 \rangle - \langle j_c\rangle^2  \right)
%\eea
%and
%\bea
%%j_q^2 = (j_E-\mu_Rj_c)^2 = j_E^2 +\mu_R^2 j_c^2 -2j_Ej_c \mu_R
%\langle j_q^2 \rangle - \langle j_q\rangle ^2 &=&
%\langle j_E^2\rangle - \langle j_E\rangle^2 
%+\mu_R^2 \left( \langle j_c^2\rangle - \langle j_c\rangle^2  \right)
%\nonumber\\
%&-&2\mu_R\left(\langle j_Ej_c \rangle - \langle j_c\rangle \langle j_E\rangle \right).
%\eea
% 
We now generalize this result for higher order ($n\geq 1$) cumulants. Since the stochastic energy current goes hand in hand with the stochastic electron flux i.e.,$I_u=\epsilon_d\,I_e$,  higher order cumulants satisfy $\langle I_u^n\rangle_c$ = $\epsilon_d^n\langle I_e^n\rangle$ and $\langle I_q^n\rangle_c = (\epsilon_d - \mu_R)^n \langle I_e^n \rangle_c$ resulting in %for $n$-th order fluctuation 
\bea
\eta^{(n)} &\equiv& \frac{\langle (-\dot{ w})^n\rangle_c}{\langle \dot q_h^n\rangle_c}
\nonumber\\
&=&
\frac{(\Delta \mu)^n}{(\epsilon_d-\mu_R)^n}= \langle\eta\rangle^n \leq \eta_C^n.
\eea
Since $\langle\eta\rangle\leq \eta_C$ we conclude that in the TC limit, $\eta^{(n)}\leq \eta_C^n$.  This result holds arbitrarily far from equilibrium.
%Moreover,  it is easy to show that high order fluctuations for efficiency behave in a similar manner and one can show that in this limit $\eta^{(n)} \leq \eta_C^n$. \cite{}  with equality obtained in the quasi-static limit.
%This is the first important result of our work.
%Next, based on simulations
%we demonstrate that this bound holds beyond the quasi-static limit.
%

%%==============================================================
% Figure S4
 \begin{figure}[h]
\centering
\includegraphics[width=0.7\textwidth]{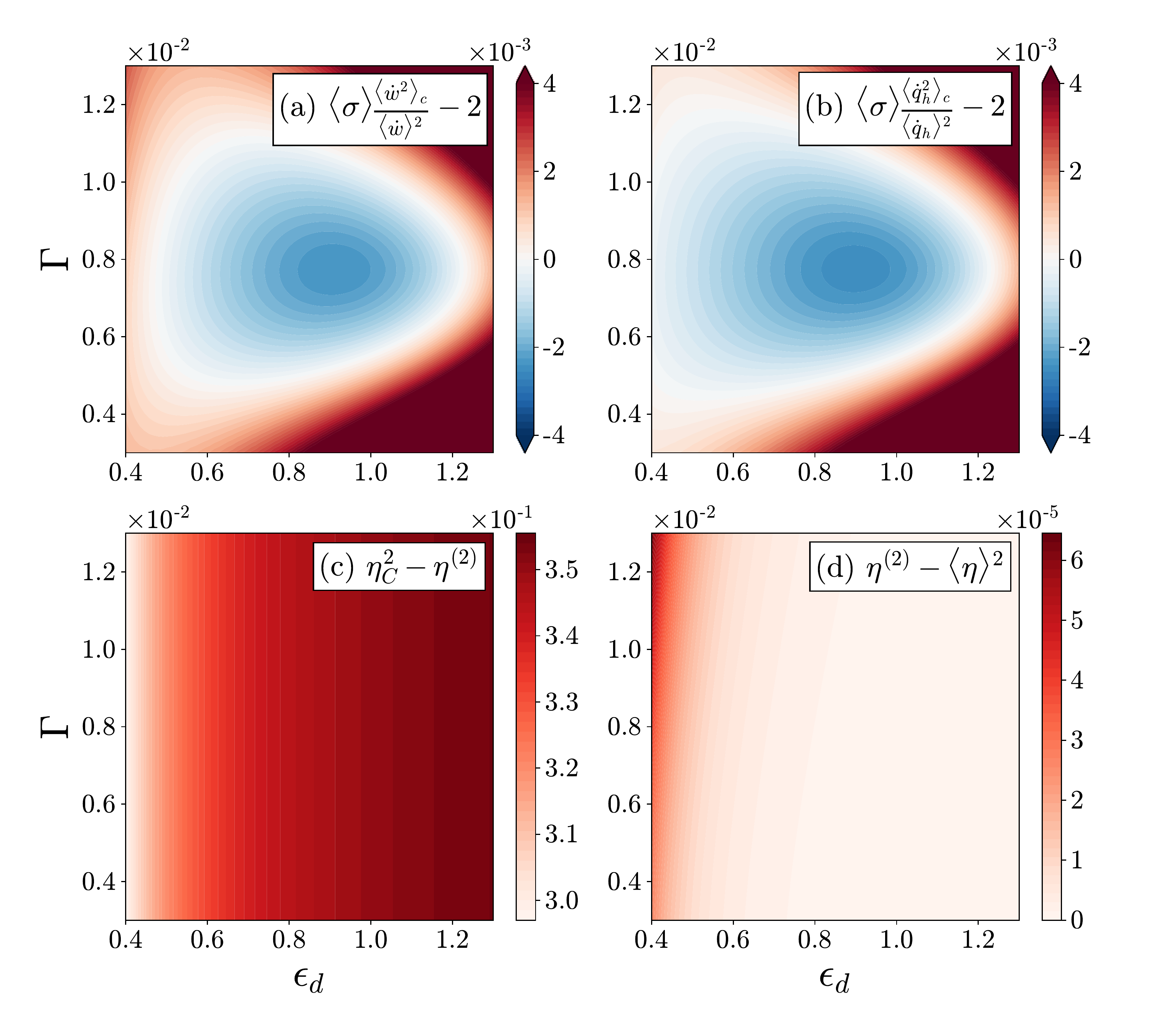}
\caption{(Color online) The TUR is violated for the work (a) and heat (b) currents in a serial double-dot thermoelectric engine. However, the upper (c) and lower (d) bounds on $\eta^{(2)}_{\textrm{eng}}$ are still satisfied. 
We use degenerate levels with intersite tunneling energy $\Omega=0.005$ and equal hybridization energies to the leads, $\Gamma$. $\beta_L=1.0$, $\beta_R = 0.4$, $\mu_L=0.1$, $\mu_R=0$.}
\label{TUR-comparison}
\end{figure}

%===========================================
\subsection{TUR violation---and the validity of our bounds}

The standard TUR, $\langle \sigma\rangle \frac{\langle I^2\rangle_c}{\langle I \rangle ^2}\leq 2$, can be violated in quantum transport junctions.  Specifically for thermoelectric junctions, 
the TUR can be violated for both work current, $\dot{w}$, and heat current, $\dot{q}_h$, in a symmetric serial double-dot thermoelectric system behaving as an engine for a limited range of values of $\epsilon_d$ and $\Gamma$
\cite{Junjie-TUR}. 
This behavior is shown in Fig.~\ref{TUR-comparison}(a)-(b).
Here, $\epsilon_d$ is the energy of the levels and $\Gamma$ is the (identical) coupling of the left and right dots to the respective metals. The intersite tunneling is denoted by $\Omega$.
However, we find that the ratio $\mathcal{Q}$ of relative fluctuations between these two quantities is still greater than 1, thus satisfying the bounds  on $\eta^{(2)}$ independently of the TUR itself, see Fig.~\ref{TUR-comparison}(c)-(d). 
We reiterate that the proof in the main text for $\mathcal{Q}\geq1$, translating to $\langle \sigma\rangle \frac{\langle \dot w^2\rangle_c}{\langle \dot w \rangle ^2} \geq\langle \sigma\rangle \frac{\langle \dot q_h^2\rangle_c}{\langle \dot q_h \rangle ^2}$
for an engine  was achieved under linear response. We point out that
Fig. ~\ref{TUR-comparison} displays the validity of this inequality outside this strict regime.
%This is reflected in the fact that the TUR violation for $\dot{q}_h$ is greater than that for $\dot{w}$, and in the fact that both the upper and lower bounds on $\eta^{(2)}_\textrm{eng}$ for this device are satisfied in the region of the violations.
% DDD: Our proof Q>1 was for linear response, so its actually not a trivial outcome. 

 %%============================================================
 
 \subsection{Optimization process for verifying the bounds}
Following the main text we expand here on the stochastic optimization process we are using to verify the bounds beyond linear response. The following procedure is implemented in JAX~\cite{JAX} and uses the ADAM algorithm~\cite{Adam}. 

We set the biases $\Delta \mu$ and $\Delta T$. We then start at an initial point in parameter space, $\left( \epsilon_d^{(i)}, \Gamma^{(i)} \right)$ and aim to minimize the following cost function
\bea
f\left( \epsilon_d, \Gamma \right) = \left( \eta^{(2)} - \langle \eta \rangle^2 \right)^2 + P\left(\langle I_e \rangle \right),
\eea
where the first term can be replaced with the equivalent difference for the upper bound, and $P\left( \epsilon_d, \Gamma \right)$ is a penalty function, which depends on the charge current's direction, and ensures the engine regime. $P\left( \epsilon_d, \Gamma \right) = 10+\frac{10^{-5}}{10^3 \textrm{ReLU}(\langle I_e \rangle)-10^{-6}}$, where ReLU is a rectifier linear unit. This specific choice of penalty function attempts to ensure a high penalty when the current is positive (outside the engine regime). It is zero for negative currents. 
Near the (engine-no engine) boundary, the currents assume very small values, which could be $\mathcal{O}(10^{-9})$. This issue lead us to set the prefactors as indicated. 

Through the optimization process for a specific initial choice $\left( \epsilon_d^{(i)}, \Gamma^{(i)} \right)$, a path of points (trajectory) is created. This path can, at times, run away from the desired engine regime. For that, we store the minimal difference, $ \eta^{(2)} - \langle \eta \rangle^2$ along each path. Since the difference is a smooth function, if it flips sign, 
we expect that at least one path should have a negative difference in the vicinity of the minima we find within the engine region. Yet, we did not encounter negative differences in the engine domain
% DDD: we define as positive current from the right side, so this doesnt agree with the sign convention, ($\langle I_e \rangle < 0 $) 
%Should be ($\langle I_e \rangle > 0 $ and $\langle I_q^R\rangle>0$)  
along any of the paths, hence, we are certain the bounds are obeyed for this choice of biases.

%====================================
\section{Single affinity limit}
% DDD I dont think it makes sense here to refer to output and input currents, but the issue is the "primary" current, and the coupled current (due to Onsager).
Here we discuss the behavior of the ${\cal Q}$ ratio in the single affinity limit. Let us first turn off the affinity corresponding to the input channel, i.e.,$A_1=0$. One then receives in the linear-response limit, 
\bea
{\cal Q}_{A_1=0} &\equiv& \frac{\langle I_1 \rangle^2}{\langle I^2_1 \rangle_c}  \,\, \frac{\langle I^2_2 \rangle_c}{\langle I_2 \rangle^2}= \frac{L_{12}^2}{L^2_{22}} \, \frac{L_{22}}{L_{11}} = \frac{L_{12}^2}{L_{22} \, L_{11}} \leq 1,
\eea
where the inequality follows from the positivity of entropy production in a spontaneous process.
%Onsager's relation. 
This result implies that the relative fluctuation or the precision of the 
current related to the applied bias ($I_2$)
%
%output current ($I_2$) 
is always upper bounded by the corresponding precision for the current of the coupled phenomena ($I_1$).
%input current ($I_1$). 
This further indicates that in the linear response limit, the TUR ratio for a %output
current under its bias, is always upper bounded by the corresponding TUR ratio for the coupled current i.e., $\langle \sigma \rangle \frac{\langle I_2^2\rangle_c}{\langle I_2 \rangle^2}\leq \,
 \langle \sigma \rangle  \frac{\langle I_1^2\rangle_c}{\langle I_1 \rangle^2}$.
Similarly, in the opposite limit, i.e., for $A_2=0$, following similar steps, it is easy to see that $Q_{A_2=0} \geq 1$ i.e., we reach the same conclusion: the current that corresponds to the applied bias is upper bounded by the  relative fluctuation of the coupled current.
%the relative fluctuation of the output current is always lower bounded by the corresponding relative fluctuation of the input current. 
%To put it simply, this results provides a simple way to control and manipulate relative fluctuations of currents by switching on and/off the thermodynamic affinities.  
% I do not think this "switching" is clear. It sounds like dynamics.

%XXXXX DO WE NEED FIGURES FOR THIS RESULT IN THE THERMOELECTRIC CASE?

%========================================
\section{Quantum Absorption Refrigerators}

We describe here the procedure for calculating heat currents and their fluctuations for multi-level QARs in the incoherent transport limit. Additional details can be found in \cite{Segal18}. More advanced simulations, taking into account quantum coherences were performed in Ref. \cite{Junjie21}.

In our setup, a multi-level system is coupled to three heat baths ($b=w,h,c$).
The objective of interest is the characteristic function, $ |Z(\chi, t)\rangle$. For a three-bath problem it
is sufficient to consider two counting parameters $\chi=(\chi_c,\chi_w)$ when studying properties in steady state, given conservation of energy. The characteristic function satisfies a first-order differential equation,
\bea \frac{ d| Z(\chi, t)\rangle}{dt}=
\hat W(\chi)|Z(\chi, t)\rangle.
\label{eq:ZW}
\eea
For the three-level model presented in the main text, Fig. 3a, the rate matrix is
\bea \hat W(\chi)=
\begin{pmatrix}
-k_{1\rightarrow 2}^c-k_{1\rightarrow 3}^h  &  k_{2\rightarrow 1}^c e^{-i\chi_c\theta_c}   & k_{3\rightarrow 1}^h  \\
k_{1\rightarrow 2}^c e^{i\chi_c \theta_c}  &   -k_{2\rightarrow 1}^c - k_{2\rightarrow 3  }^w & k_{3\to 2}^w e^{-i\chi_w\theta_w}\\
k_{1\rightarrow 3}^h   &   k_{2\rightarrow 3}^w e^{i\chi_w\theta_w} &  -k_{3\rightarrow 2 }^w -k_{3\to 1}^h\\
\end{pmatrix}
\nonumber\\
\label{eq:M}
\eea
Here, we count the three levels (bottom to top) by 1,2,3,
with energies $\epsilon_{1,2,3}=(0, \theta_c, \theta_w)$.
Assuming that the baths comprise collections of harmonic oscillators, which are coupled to the quantum system through the displacements of the oscillators, the rate constant due to the $b$ bath is given by the product
\bea
k_{i\to j}^b = J_b(\epsilon_j-\epsilon_i)n_b(\epsilon_j-\epsilon_i). 
\eea
Here, $n_b(\Delta)=\left(e^{\beta_b\Delta}-1\right)^{-1}$ is the Bose-Einstein distribution function and
$J_b(\Delta)=\gamma_b\Delta e^{-|\Delta|/\Lambda}$ is the spectral density function of the $b$ bath.
For simplicity, we assume Ohmic spectral functions and take the cutoff frequency to be large, $\Lambda\gg \theta_{c,w},1/\beta$.
 
For the four-level model as presented in Fig. 3b of the main text, the rate matrix is
\begin{widetext}
\begin{equation}
%\mathcal{\sum_{\alpha}D_{\alpha}}=
\hat W(\chi)=
\begin{pmatrix}
-k_{1\to 2}^{c} -k_{1\to 3}^{c} -k_{1\to 4}^h & k_{2\to 1}^{c}e^{-i\chi_c(\theta_c-g)} & k_{3\to 1}^{c} e^{-i\chi_c(\theta_c+g)}& k_{4\to 1}^h\\
k_{1\to 2}^{c}e^{i\chi_c(\theta_c-g)} &-k_{2\to 1}^{c} -k_{2\to 4}^w&0 & k_{4\to 2}^we^{-i\chi_w(\theta_w+g)} \\
k_{1\to 3}^{c} e^{i\chi_c(\theta_c+g)} &0& -k_{3\to 1}^{c} -k_{3\to 4}^w & k_{4\to 3}^w e^{-i\chi_w(\theta_w-g)}\\
k_{1\to 4}^h &  k_{2\to 4}^w e^{i\chi_w(\theta_w+g)}   &  k_{3\to 4}^w e^{i\chi_w(\theta_w-g)}& -k_{4\to 1}^h - k_{4\to 2}^w -k_{4\to 3}^w 
\end{pmatrix}
\end{equation}
\end{widetext}
We count the levels 1 to 4 from bottom to top,
with energies $\epsilon_{1,2,3,4}=(0, \theta_c-g, \theta_c+g, \theta_c+\theta_w)$.

The cumulant generating unction (CGF) is given in terms of the characteristic function,
\bea
G(\chi) =  \lim_{t \to \infty} \ \frac{1}{t}\ln \langle I| Z(\chi,t) \rangle,
\eea
with $\langle I|$ a unit vector
($\langle I= \langle 1 1 1 1|$ for the 4-level QAR).
To obtain the CGF,  we diagonalize $\hat W(\chi)$ and select the eigenvalue that dictates the long-time dynamics---this is the eigenvalue with the smallest magnitude for its real part. 
Once we reach the CGF,  the currents  and the second cumulants are given by ($b=c,w$)
\bea
\langle I_{b} \rangle&=& \frac{\partial G}{\partial (i\chi_{b})}\Big|_{\chi=0},
\nonumber\\
\langle I_{b}^2\rangle_c &=&  \frac{\partial^2 G}{\partial (i\chi_{b})^2}\Big|_{\chi=0}.
\eea
This procedure can be done analytically for the three-level QAR.
In the main text we obtain the current and its fluctuations numerically: we diagonalize numerically the rate matrix $\hat W(\chi)$ for a range of $\chi$. The cumulants are then obtained using finite differences.

%==========================================

\end{widetext}
%\end{document} 

%% file: bound_avg.bbl
\begin{thebibliography}{99}
\bibitem{comment}
The chemical potential difference $\Delta \mu$ is attached here to the affinity, with the  charge current as the corresponding current.
If the output current is the power, $-\langle \dot w\rangle = \Delta \mu\langle I_e\rangle $, then the corresponding affinity should be stripped of $\Delta \mu$.
\end{thebibliography}
